\documentclass[
 reprint,
 floatfix,
 amsmath,amssymb,
 superscriptaddress,
 aps,
 prapplied,
 nofootinbib,
]{revtex4-2}

\usepackage{siunitx}
\usepackage{graphicx}
\usepackage{dcolumn}
\usepackage{bm}
\usepackage{physics2}
\usephysicsmodule{ab}
\usephysicsmodule{op.legacy}
\usephysicsmodule{braket}
\usephysicsmodule{nabla.legacy}
\usepackage{derivative}
\usepackage{mathtools}

\usepackage{xcolor}

\usepackage{appendix}
\usepackage[colorlinks,linkcolor=blue,urlcolor=blue, citecolor=blue]{hyperref}
\usepackage[T1]{fontenc}
\usepackage{comment}
\usepackage{amsfonts}
\usepackage{multirow}
\usepackage{makecell}

\newcommand{\mqty}[1]{\ab(\begin{matrix}#1\end{matrix})}
\newcommand{\dd}[1]{\odif{#1}\,}

\usepackage{newtxtext}
\usepackage{newtxmath}

\begin{document}
\title{Passive quantum interconnects:\\ multiplexed remote entanglement generation with cavity-assisted photon scattering}
\author{Seigo Kikura}
\email{seigo.kikura@nano-qt.com}
\affiliation{Nanofiber Quantum Technologies, Inc. (NanoQT), 1-22-3 Nishiwaseda, Shinjuku-ku, Tokyo 169-0051, Japan}

\author{Kazufumi Tanji}
\affiliation{Nanofiber Quantum Technologies, Inc. (NanoQT), 1-22-3 Nishiwaseda, Shinjuku-ku, Tokyo 169-0051, Japan}

\author{Akihisa Goban}
\email{akihisa.goban@nano-qt.com}
\affiliation{Nanofiber Quantum Technologies, Inc. (NanoQT), 1-22-3 Nishiwaseda, Shinjuku-ku, Tokyo 169-0051, Japan}

\author{Shinichi Sunami}
\email{shinichi.sunami@nano-qt.com}
\affiliation{Nanofiber Quantum Technologies, Inc. (NanoQT), 1-22-3 Nishiwaseda, Shinjuku-ku, Tokyo 169-0051, Japan}
\affiliation{Clarendon Laboratory, University of Oxford, Oxford OX1 3PU, United Kingdom}

\begin{abstract}

We propose a time- and wavelength-multiplexed remote atom-atom entanglement generation protocol based on cavity-assisted photon scattering (CAPS).
This is designed to achieve a high rate and high fidelity with robustness to operational imperfections, parameter fluctuations, and auxiliary time costs, such as percent-level photon impurity, timing and cavity parameter jitter, and atom shuttling time costs.
We benchmark this protocol using comprehensive analytical and numerical modeling of the atom-cavity dynamics, including state-dependent pulse delay effects, photon temporal impurity, atom-cavity system parameter fluctuations, and crosstalk among atoms through a shared cavity mode.
With realistic atom-cavity system performance, we predict \qty{2e5}{s^{-1}} successful atom-atom Bell pair generation even without in-cavity qubit reset, substantially enhanced from two-photon-interference-based protocols, at a predicted heralded fidelity of 0.999.

\end{abstract}

\maketitle

\section{Introduction}
Construction of large-scale fault-tolerant quantum computers is one of the central goals of quantum technologies.
The required number of physical qubits for various classically intractable problems is estimated to exceed millions, due to the overhead associated with quantum error correction~\cite{Gidney2021,Beverland2022}.
Building such systems within a single monolithic device presents substantial technical and architectural challenges.
Modular architectures that interconnect smaller quantum processors via optical links offer a promising and practical solution~\cite{Monroe2014, Covey2023, Sunami2025}.
Beyond scalability, high-performance optical interconnects enable a broad range of applications such as blind quantum computing~\cite{Fitzsimons2017}, long-baseline quantum sensing~\cite{Gottesman2012, Khabiboulline2019}, and long-distance quantum communication~\cite{Azuma2023}.
The key performance metrics of such interconnects are the fidelity and the rate of remote entangled qubit pair generation.
High fidelity reduces the large overhead for entanglement distillation required for fault-tolerant operation~\cite{Pattison2024}, while a high rate ensures sufficient bandwidth for inter-module gate execution~\cite{Sunami2025}.

For atomic qubit platforms such as neutral atoms and trapped ions, conventional photon-emission-based protocols proceed with an atom-state-dependent emission of photons into separate modes, such as polarization, time-bin, and frequency modes, which are detected after the two-photon interference at beamsplitters, for a heralded generation of maximally entangled states of atomic qubits with a practical upper bound of 50\% success probability~\cite{Duan2003,Beukers2024}.
Both high fidelity and rate are expected with the aid of optical cavities~\cite{Li2024,Sinclair2025,Sunami2025}; however, this requires fine-tuning of the atom-photon coupling strengths of the two parties~\cite{Li2024, Kikura2025}, careful management of the emission-induced recoil effect~\cite{Kikura2025}, fast, high-power excitation laser pulses with stringent inter-module synchronization requirements~\cite{Li2024, Kikura2025}, and many rounds of entanglement trials~\cite{Huie2021, Li2024, Sunami2025}.

To address these challenges, an attractive alternative for remote entanglement generation is based on the reflection of light pulses from the one-sided cavity for a controlled phase flip gate between atomic and photonic qubits~\cite{Duan2004, Reiserer2014, Tiecke2014, Volz2014}, which we call the cavity-assisted photon scattering (CAPS) protocol.
This has several critical advantages, such as robustness against various imperfections, including mismatches and fluctuations in atom-cavity parameters across the network, operation without the need for fast atom excitation pulses, a higher success probability, and the absence of the need for inter-module synchronization~\cite{Duan2005, Lin2006}.
The flexibility of the CAPS gate also allows for a wider variety of applications, including heralded memory loading, photon-photon gates, nondestructive photon detection, and remote atom-atom gates~\cite{Su2012, Kalb2015, Hacker2016, Distante2021, Welte2018, Knaut2024}.

Despite these advantages, the CAPS-based remote entanglement generation protocol is considered, within the conventional framework, to demand optical cavities of exceptionally high quality for high-fidelity operations~\cite{Goto2010, Asaoka2021, Asaoka2025}.
Furthermore, the imperfections of the single photon used to mediate the entanglement, which are inherent in realistic implementations, are not well investigated, leaving the evaluation under practical settings open.
The fidelity of CAPS-based networking operations is also currently known to degrade rapidly with shorter optical pulses, resulting in a fundamental rate-fidelity tradeoff with unfavorable scaling~\cite{Duan2004,Utsugi2025}.
It is also unclear whether the time-multiplexed operations with many atoms, known to enhance the entanglement generation rate by orders of magnitude compared to that of single-atom network nodes~\cite{Huie2021,Li2024,Sunami2025}, will be realistic for the CAPS-based approach.

To resolve these issues, we develop a comprehensive theoretical framework to design and evaluate high-rate, high-fidelity CAPS-based atom-photon interactions.
This framework incorporates a wide variety of imperfections, such as losses and group delay of photon wavepackets upon reflection from the cavity, the effect of mixed temporal modes of the photon (photon impurity), cavity and photon spectral shifts, finite spectral width of the input photon, as well as cavity-mode-induced crosstalk in the case of multiple atoms coupled to the cavity for time-multiplexed operations; so far, no such theoretical framework has been available, and we further incorporate several critical challenges that were previously overlooked by developing a comprehensive evaluation procedure for CAPS-based multi-node networking.
This allows us to propose concrete protocols to mitigate these error sources and, furthermore, to identify novel CAPS-based protocols that overcome limitations in practical implementation and reduce the required hardware (for example, by eliminating the need for an independent photon source), as well as to demonstrate their performance in realistic settings.

Consequently, our work simultaneously enhances the performance of cavity-based quantum interconnects while reducing the required hardware performance compared to commonly employed two-photon-interference protocols.
By establishing asynchronous, `passive' quantum interconnects with tolerance for varying device parameters and a wide range of imperfections, our proposed protocols enable the scalable implementation of large-scale quantum networks.

\begin{figure}[t]
    \centering
    \includegraphics[width=0.95\linewidth]{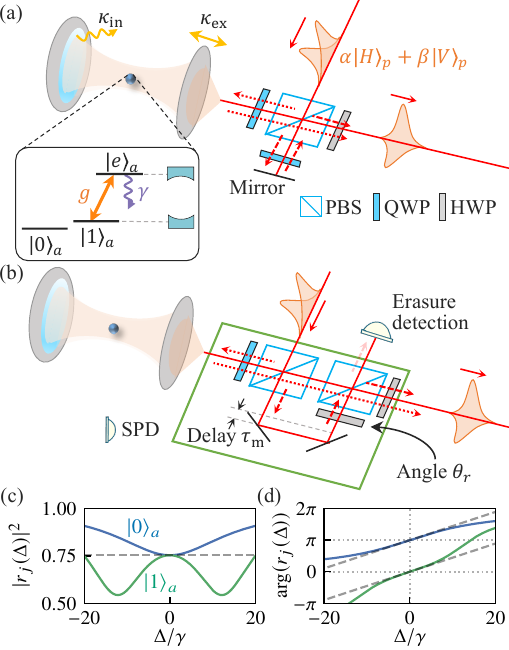}
    \caption{High-fidelity CAPS gate.
    (a) CAPS protocol.
    A $\ket{1}_a \leftrightarrow \ket{e}_a$ transition of a three-level atom is resonantly coupled to a one-sided optical cavity with coupling rate $g$.
    The cavity interfaces with the propagating mode at rate $\kappa_\text{ex}$ with internal loss at rate $\kappa_\text{in}$, while the decay rate of the excited state of the atom $\ket{e}_a$ is $\gamma$.
    An incoming polarization-encoded photonic qubit (top) is split at a first polarizing beamsplitter (PBS): the initially $V$-polarized component is routed to a one-sided cavity through a quarter-wave plate (QWP), reflected off from the cavity and back to the device towards the output port (dotted arrows), while the $H$-polarized component first transmits through the PBS, QWP and a mirror, before reflected from the PBS to be recombined with the other polarization mode (dashed arrows).
    Overall, this protocol implements a CZ gate between the atomic qubit (encoded in $\ket{0}_a, \ket{1}_a$ basis) and a photonic qubit~\cite{Duan2004, Zhang2016}.
    (b) High-fidelity CAPS gate implemented with a modified optical layout (green rectangle).
    A controllable photon loss is induced for the initially $H$-polarized mode by a half-wave plate (HWP); a single-photon detector (SPD) heralds gate failure without disturbing the atomic qubit (see text).
    The calibrated path delay $\tau_\text{m}$ is introduced to cancel the effect of pulse delay arising from the cavity dispersion.
    (c) Cavity reflectivity $|r_j(\Delta)|^2$ as a function of the detuning $\Delta/\gamma$ for atomic states $\ket{j}_a=\ket{0}_a$ (blue) and $\ket{1}_a$ (green) with an optimized cavity-QED system that satisfies Eqs.~(\ref{eq:kappa_ex_opt},\ref{eq:condition_for_kappa_in_gamma}) with $C_\text{in} = 100$.
    Both reflectivities match at $\Delta = 0$ as $(r^\text{opt})^2$ (dashed line).
    (d) Phase shift upon cavity reflection, $\arg(r_j(\Delta))$.
    At $\Delta=0$, the phase difference is exactly $\pi$, and both slopes match as $\gamma\tau_\text{m}$ (dashed lines).
    }
    \label{fig1}
\end{figure}

\section{High-fidelity cavity-assisted photon scattering}
In this section, we first review the conventional CAPS protocol and integrate several recent advances for improved fidelity of CAPS operation into a common framework, along with an experimentally implementable optical layout for error mitigation techniques.
Figure~\ref{fig1}(a) illustrates the CAPS protocol in a conventional setting.
An incoming polarization-encoded photonic qubit, $\ket{\psi}_p = \alpha \ket{H}_p + \beta \ket{V}_p$, is sent to a polarizing beamsplitter (PBS) that first splits the two polarization components: the $V$-polarized component is reflected off the PBS to be routed to the cavity mirror (dotted arrows).
After the reflection from the cavity and passing through the PBS again, the $V$-polarized component is recombined with the $H$-polarized component that reflects off the PBS and is routed back by standard mirrors (dashed arrows).
Inside the one-sided cavity, a three-level atom with internal states, $\ket{0}_a, \ket{1}_a$ and $\ket{e}_a$, is coupled to the cavity mode through a $\ket{1}_a \leftrightarrow \ket{e}_a$ transition that is resonant with the cavity.
When the cavity, the photon, and the atomic transition are all on resonance, the photon reflecting off the cavity mirror acquires a $\pi$ phase shift if the atom is in $\ket{0}_a$.
Combined with the optical layout illustrated in Fig.~\ref{fig1}(a), a controlled-phase (CZ) gate between the atomic and photonic qubits is possible in a passive manner with no synchronization required, which we call the CAPS gate~\cite{Duan2004}.
Henceforth, we may relabel the photonic basis states as $\ket{0}_p\equiv\ket{H}_p$ and $\ket{1}_p\equiv\ket{V}_p$ unless stated otherwise.

Below, we propose a high-fidelity CAPS protocol by identifying and canceling several leading-order error sources, using a modified layout [Fig.~\ref{fig1}(b)].
The atomic-state dependent reflection functions of resonantly coupled atom-cavity systems are given by~\cite{Cohen2018, Raymer2024, Utsugi2025}
\begin{equation} \label{eq:freq-dependent-reflectance}
    \begin{aligned}
        {r}_0(\Delta) =& \frac{-\kappa_\text{ex} + \kappa_\text{in} -i\Delta}{\kappa_\text{ex}+\kappa_\text{in} - i\Delta}, \\
        {r}_1(\Delta) =& \frac{(-\kappa_\text{ex} + \kappa_\text{in} -i \Delta)(\gamma-i\Delta) + g^2}{(\kappa_\text{ex}+\kappa_\text{in} - i\Delta)(\gamma-i\Delta) + g^2},
    \end{aligned}
\end{equation}
where $\Delta$ is the detuning of the incident-photon frequency from the atomic transition.
See Fig.~\ref{fig1} for the definition of the other parameters.
Here, we analytically investigate the zeroth- and first-order errors in $\Delta$, which constitute the dominant noise sources in the CAPS gate, and their mitigation by tuning the system parameters.
Higher-order effects are evaluated by numerical simulations presented later.

\begin{table*}[t]
    \caption{Summary of protocols for generating remote entanglement on two qubits via photons. The symbols $P_\text{gen}$ and $P_\text{gen}'$ denote the emission probability of photon sources and that of atom-photon entanglement sources (ESs) in a single trial, respectively.}
    \small
    \begin{ruledtabular}
        \begin{tabular}{llll}
            Protocol  & Success probability & Heralded fidelity & Hardware requirements \\
            \hline
            Remote CZ gate & $P_\text{gen}P_\text{rg} \simeq (1-\sqrt{2/C_\text{in}})P_\text{gen}P_\text{CAPS}^\text{opt}$\, [Eq.~\eqref{eq:def_P_rg}] & see Ref.~\cite{Goto2010, Asaoka2021}  & Photon source, two atom-cavity systems \\
            Sequential CAPS [Fig.~\ref{fig2}] & $P_\text{cc} \simeq P_\text{gen} P_\text{CAPS}^\text{opt}$\, [Eq.~\eqref{eq:def_P_cc}] & $F_\text{cc}$ [Eq.~\eqref{eq:F_cc}] & Photon source, two atom-cavity systems  \\
            Emission-CAPS [Fig.~\ref{fig3}] & $P_\text{ec} \simeq P_\text{gen}' P_\text{CAPS}^\text{opt}$ [Eq.~\eqref{eq:P_ec^(j)}] & $F_\text{ec}$ [Eq.~\eqref{eq:F_ec}] & Two atom-cavity systems \\
            Two-photon interference [Fig.~\ref{fig3}] & $P_\text{ee}\simeq (P_\text{gen}')^2/2$ [Eq.~\eqref{eq:P_ee}] & $F_\text{ee}$ [Eq.~\eqref{eq:F_ee}] & Two atom-cavity systems
        \end{tabular}
    \end{ruledtabular}
    \label{tab:remote_entanglement_generation_protocols}
\end{table*}
The zeroth-order error, independent of the photon temporal envelope, represents the total photon loss and the unbalanced atomic-state-dependent loss, i.e.,~$\vab{r_j(0)} < 1\, (j\in\{0,1\})$ and $\vab{r_0(0)} \neq \vab{r_1(0)}$.
To eliminate this error, one can tune the output coupling strength $\kappa_\text{ex}$, which is implementable by adjusting the (effective) cavity mirror transmittance, to~\cite{Goto2010}
\begin{equation} \label{eq:kappa_ex_opt}
    \kappa_{\text{ex}}^{\text{opt}} = \kappa_{\text{in}}\sqrt{1+2C_{\text{in}}},
\end{equation}
which balances the losses, as shown in Fig.~\ref{fig1}(c).
In particular, on resonance,
\begin{equation} \label{eq:opt_r}
    -r_{0}(0) = r_{1}(0) = 1-\frac{2}{1+\sqrt{1+2C_{\text{in}}}}\;\eqqcolon\; r^{\mathrm{opt}}.
\end{equation}
Here, $C_{\text{in}} = g^{2}/(2\kappa_{\text{in}}\gamma)$ is the internal cooperativity, quantifying the internal-loss-limited quality of the atom-cavity system~\cite{Goto2019}.
With the adjustment of the reflection amplitude $r_\text{m}$ for $\ket{0}_p$, where necessary, this enables the cancellation of the zeroth-order error.

The first-order error arises from the atomic-qubit-dependent pulse delay, resulting in an incomplete overlap of the reflected photon and inducing a substantial error in CAPS operations with realistic finite-duration pulses~\cite{Utsugi2025} (see Appendix~\ref{s_sec:optimize_caps_gate} and Fig.~\ref{fig7}).
Such an error can also be canceled by tuning the coupling rate as~\cite{Hastrup2022, Utsugi2025}
\begin{equation} \label{eq:kappa_ex_delay}
    \kappa_{\text{ex}}^{\text{delay}}
    =\sqrt{\kappa_{\mathrm{in}}^{2}+2\gamma\kappa_{\mathrm{in}}+g^{2}},
\end{equation}
resulting in an atomic-state-independent pulse delay [see Fig.~\ref{fig1}(d)]
\begin{equation} \label{eq:tau_m}
    \tau_\text{delay} = \frac{2\kappa_\text{ex}}{\kappa_\text{ex}^2-\kappa_\text{in}^2}.
\end{equation}
To ensure that the $H$- and $V$-polarized photons also overlap, we further introduce the calibrated delay of the $H$-polarized photon, $\tau_\text{m} = \tau_\text{delay}$, using the optical layout as shown in Fig.~\ref{fig1}(b).

It is possible to meet the two independent requirements for $\kappa_\mathrm{ex}$ in Eqs.~(\ref{eq:kappa_ex_opt},\ref{eq:kappa_ex_delay}), i.e., $\kappa_{\text{ex}}^{\text{opt}} = \kappa_{\text{ex}}^{\text{delay}}$, by tuning the ratio between the atomic decay and cavity internal-loss rates,
\begin{equation} \label{eq:condition_for_kappa_in_gamma}
    \frac{\kappa_\text{in}}{\gamma} = \frac{1+C_\text{in}}{C_\text{in}},
\end{equation}
which is possible by an appropriate design of the resonator length $L_\mathrm{cav}$~\cite{Utsugi2025};
this is because, while $\kappa_\mathrm{in}$ strongly depends on the length ($\kappa_\mathrm{in} \propto 1/L_\mathrm{cav}$), $\gamma$ remains constant for varying cavity lengths, and so does $C_\mathrm{in}$ for optical cavity designs with negligible propagation loss~\cite{Shadmany2025,Sunami2025, Horikawa2025,Grinkemeyer2025}.
The optimal cavity length is given by (see Appendix~\ref{s_sec:optimize_caps_gate}, as well as Ref.~\cite{Utsugi2025})
\begin{equation} \label{eq:L_cav^opt}
    L_\text{cav}^\text{opt} = \frac{1}{1+C_\text{in}}\frac{\sigma_0}{A_\text{eff}}\frac{c}{2\gamma},
\end{equation}
where $\sigma_{0}$ is the resonant absorption cross-section, $A_{\text{eff}}$ is the effective mode area, and $c$ is the speed of light.
The resulting optimal values are typically on the order of centimeters to several tens of centimeters, representing a typical operating regime for several cavity implementations such as bow-tie cavities~\cite{Chen2022bowtiecav, Peters2024cavityrealtime}, Fabry-P\'erot cavities~\cite{Kroeze2023confocalmicroscope, Shadmany2025}, and nanofiber cavities~\cite{Horikawa2024, Horikawa2025}, many of which feature postfabrication length tuning capabilities.
This allows several leading-order errors in the CAPS gate to be canceled, resulting in substantial gate fidelity improvement even for realistic cavity qualities and pulse durations (see Appendix~\ref{s_sec:optimize_caps_gate} for a detailed performance analysis).
To see this in a more relevant setting, we present a performance analysis of remote atom-atom entanglement generation in the following sections; the protocols we discuss and develop are summarized in Table~\ref{tab:remote_entanglement_generation_protocols}, for reference.

\begin{figure*}[t]
    \centering
    \includegraphics[width=0.9\linewidth]{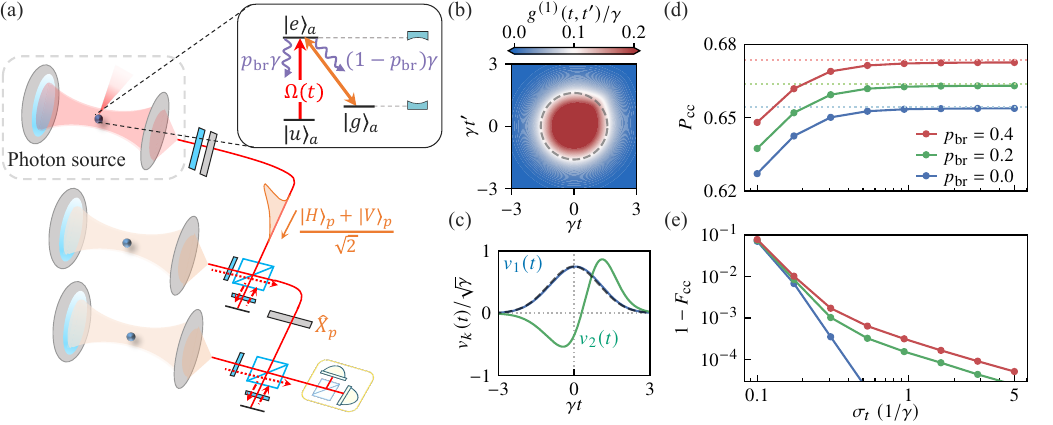}
    \caption{Remote entanglement generation with CAPS gates.
    (a) Schematic of the CAPS-based remote entanglement generation with cavity-QED-based photon source.
    An atom-cavity system provides a single photon to be routed to other cavities for mediating atom-atom entanglement.
    The atom coupled to the source cavity has three levels, $\ket{u}_a, \ket{e}_a$ and $\ket{g}_a$, where excitation laser is used to excite to $\ket{e}_a$ from which the atom decays to $\ket{u}_a$, or $\ket{g}_a$, with branching ratio $p_\mathrm{br}$ where $p_\mathrm{br}>0$ results in reexcitation-induced impurity of the photon.
    (b) Autocorrelation function of the emitted photon, where the parameters for the source system are $C_\text{in} = 10$ and $p_\text{br} = 0.5$, and the Rabi frequency is set to generate the Gaussian wavepacket photon with $\sigma_t = 1/\gamma$.
    The dashed line is a guide to the eye to highlight the small tail at the top right region.
    (c) Two primary eigenmodes $v_1(t)$ and $v_2(t)$ with the corresponding eigenvalues $\lambda_1 = 0.68$ and $\lambda_2 = 0.025$, respectively $(P_\text{gen}=\sum_k \lambda_k = 0.72)$.
    The first mode closely matches the desired Gaussian function (dashed line), while the second exhibits a significant deviation.
    (d) Success probability of the remote entanglement generation based on sequential CAPS gates incorporating the source imperfection, where $(g,\gamma, \kappa_\text{ex}, \kappa_\text{in})$ characterize three cavity-QED systems with $C_\text{in} = 100$.
    The dotted lines represent the analytical upper bound $\bar{P}_\text{gen} P_\text{CAPS}^\text{opt}$.
    (e) Infidelity of the generated Bell pairs.
    Larger source imperfection, characterized by $p_\text{br}$, degrades generated Bell states;
    increasing $\sigma_t$ suppresses the infidelity below $10^{-4}$ even for high $p_\mathrm{br}$, leading to a tradeoff between fidelity and success rate $\propto P_\mathrm{cc}/\sigma_t$.
    }
    \label{fig2}
\end{figure*}

\section{Remote atom-atom entanglement generation via sequential CAPS gates}

Having established an optimized CAPS gate primitive under realistic imperfections, we next examine the impact of these errors at the protocol level by analyzing the remote entanglement generation via sequential CAPS gates~\cite{Duan2005} [Fig.~\ref{fig2}(a)].
We first assume that we have access to a perfect single-photon source and use this to perform remote entanglement generation of atomic qubits in two cavity-QED systems (Alice and Bob).
We later extend the analysis to a more realistic assumption of an imperfect photon source, such as atom-cavity systems, as illustrated.

Conventionally, remote atom-atom operation via CAPS is designed for a conditional remote CZ gate:
this is achieved by the successive reflection of an ancilla photon from two cavity systems, with a HWP in the middle for the Hadamard gate $\hat{H}_p$;
photon detection heralds the successful execution of the remote atom-atom controlled phase (CZ) gate~\cite{Duan2005, Goto2010, Asaoka2021}.
If the two parties have the same atom-cavity systems with internal cooperativity $C_\text{in}$, the maximum success probability is given by~\cite{Goto2010}
\begin{equation} \label{eq:def_P_rg}
    P_\text{rg} = (r^\text{opt})^2 \ab[1- \frac{(C_\text{in}-1)\sqrt{1+2C_\text{in}}}{C_\text{in}^2}],
\end{equation}
in the long-pulse limit, that is, in the limit of infinitely narrow spectral width of the photon.

In contrast, in most distributed operations, a simplified heralded generation of maximally entangled qubit pairs is often sufficient as a \textit{resource state} to perform remote operations, such as teleported remote CNOT gates~\cite{Ramette2024, Main2025, Sunami2025}.
For this, the HWP between the two cavities can be configured at an angle of $\pi/4$, i.e., we replace the photonic Hadamard gate $\hat{H}_p$ with the bit-flip gate $\hat{X}_p$, improving the success probability by negating the additional loss factor in $P_\text{rg}$, as we describe below.
Here, Alice (A) and Bob (B) each prepare the atomic qubits in $\ket{+}^{\text{A(B)}} = (\ket{0}^{\text{A(B)}}+\ket{1}^{\text{A(B)}})/\sqrt{2}$, and the photon is in $\ket{+}_p$.
For a sufficiently long pulse such that the reflection function can be approximated by its resonant amplitude $r_i(\Delta) \simeq r_i(0)$, the total system evolves as follows:
\begin{equation} \label{eq:sequential_caps_state_evolution}
    \begin{aligned}
        & \ket{+}_p \ket{+}^{\text{A}}\ket{+}^{\text{B}} \\
        \xrightarrow{\text{first reflection}} & \frac{2\ket{0}_p\ket{+}^\text{A} + r^\text{opt} \ket{1}_p (-\ket{0}^\text{A} + \ket{1}^\text{A})}{2}\ket{+}^\text{B} \\
        \xrightarrow{\hat{X}_p} & \frac{2\ket{1}_p\ket{+}^\text{A} + r^\text{opt} \ket{0}_p (-\ket{0}^\text{A} + \ket{1}^\text{A})}{2}\ket{+}^\text{B} \\
        \xrightarrow{\text{second reflection}} & -\frac{r^{\text{opt}}}{\sqrt{2}} (\ket{+}_p \ket*{\Phi^-}^{\text{AB}} + \ket{-}_p \ket*{\Psi^-}^{\text{AB}}),
    \end{aligned}
\end{equation}
where $\ket*{\Phi^\pm} = (\ket{0}\ket{0} \pm \ket{1}\ket{1})/\sqrt{2}$ and $\ket*{\Psi^\pm} = (\ket{0}\ket{1} \pm \ket{1}\ket{0})/\sqrt{2}$ represent the Bell states.
Here, while the photon is routed to two cavity systems, each polarization component of the photon reflects off the cavity only once.
Photon measurement in the $X$ basis projects the remote two-qubit state onto the Bell states with a probability of
\begin{equation}
    P_\text{CAPS}^\text{opt} = (r^\text{opt})^2,
\end{equation}
improved from the conventional protocol with $P_\text{rg}$ in Eq.~\eqref{eq:def_P_rg}.

\subsection{Finite spectral width of the ancilla photon}
Beyond the idealized long-pulse limit, we incorporate the effect of a finite spectral width of the photon by considering a pure-state single photon with a frequency spectrum $f(\Delta)$, which results in a photon emission probability of $P_\text{gen} = \int \dd{\Delta} |f(\Delta)|^2$.
Following the consecutive reflections from two cavities, the click of detector $j = 0(1)$ for $\ket{+(-)}_p$ heralds the successful generation of the atom-atom entangled state,
\begin{equation}
    \hat{\rho}_\text{cc}^{(j)} = \frac{1}{P^{(j)}_\text{cc}} \int \dd{\Delta} \frac{\vab{f(\Delta)}^2}{2} \ketbra*{\Upsilon^{(j)}(\Delta)}{\Upsilon^{(j)}(\Delta)},
\end{equation}
with the detection probability
\begin{equation} \label{eq:def_P_cc}
    P^{(j)}_\text{cc} = \int \dd{\Delta} \frac{\vab{f(\Delta)}^2}{2} \braket*{\Upsilon^{(j)}(\Delta)}{\Upsilon^{(j)}(\Delta)},
\end{equation}
where $\ket*{\Upsilon^{(j)}(\Delta)}$ is the (unnormalized) two-qubit state conditioned on photon detection in frequency $\Delta$ (see Appendix~\ref{s_sec:type-II-networking_with_photon} for the explicit form).
Thus, the overall success probability is $P_\text{cc} = \sum_{j=0,1}P^{(j)}_\text{cc}$, and the conditional average fidelity is given by
\begin{equation} \label{eq:F_cc}
    F_\text{cc} = \frac{P^{(0)}_\text{cc} \braket*[3]{\Phi^{-}}{\hat{\rho}^{(0)}_\text{cc}}{\Phi^{-}} + P^{(1)}_\text{cc} \braket*[3]{\Psi^{-}}{\hat{\rho}^{(1)}_\text{cc}}{\Psi^{-}}}{P^{(0)}_\text{cc}+ P^{(1)}_\text{cc} }.
\end{equation}

\subsection{Robustness against nonidentical systems}
In a practical situation, Alice and Bob may have atom-cavity systems that have different performances, which are characterized by the reflection functions $r^{\text{A}}_i(\Delta)$ and $r^{\text{B}}_i(\Delta)$.
Even in this case, once Alice and Bob independently calibrate their local delay lines $\tau_\text{m}^\text{q}  \,(\text{q}\in \{\text{A},\text{B}\})$ to be the same with the cavity-induced pulse delay in Eq.~(\ref{eq:tau_m}) and further adjust mirror-path amplitudes $r^{\text{q}}_\text{m} \leq 1$ by tuning the angle $\theta_r$ of HWP [see Fig.~\ref{fig1}(b)], our protocol reproduces the perfect Bell-state projection to leading order in $\Delta$; the projected atomic state is given by
\begin{equation}
    \begin{aligned}
         \ket*{\Upsilon^{(j)}(\Delta)} =& \frac{r^\text{opt,A}r_\text{m}^\text{B} + (-1)^jr_\text{m}^\text{A}r^\text{opt,B}}{2} \ket{\Phi^-} \\
        & +\frac{r^\text{opt,A}r_\text{m}^\text{B} - (-1)^jr_\text{m}^\text{A}r^\text{opt,B}}{2}\ket{\Psi^-} + \mathcal{O}(\Delta^2).
    \end{aligned}
\end{equation}
This shows that the tuning of the mirror-path amplitudes to satisfy $r^{\text{opt,A}}r^{\text{B}}_\text{m} = r^{\text{A}}_\text{m}r^{\text{opt,B}} = \min(r^{\text{opt,A}}, r^{\text{opt,B}})$ recovers the ideal Bell states at a success probability of $[\min(r^{\text{opt,A}}, r^{\text{opt,B}})]^2$.

\subsection{Imperfect-purity photons}
\label{sec:imperfect-purity}

In a realistic single-photon source, such as quantum dots, atomic emitters, and spontaneous parametric downconversion with photon detection~\cite{Eisaman2011}, the emitted photon is better described by a mixed quantum state.
For a given propagation mode, a quantum state $\hat{\varrho}$ containing at most one photon is characterized by the temporal autocorrelation function~\cite{Fabre2020} (see Appendix~\ref{s_sec:photon_generation} for more details),
\begin{equation} \label{eq:def_g1}
    g^{(1)}(t,t^\prime) = \Tr[\hat{a}^\dagger(t) \hat{a}(t^\prime)\hat{\varrho}],
\end{equation}
where $\hat{a}(t)$ is the instantaneous annihilation operator of the propagating mode, which satisfies $[\hat{a}(t), \hat{a}^\dagger(t^\prime)] = \delta(t-t^\prime)$.
The eigenmode decomposition
\begin{equation} \label{eq:g1-decomp}
    g^{(1)}(t, t^\prime) =\sum_k \lambda_k v_k^\ast(t) v_k(t^\prime),
\end{equation}
provides information about how the photon population is distributed among the mode basis $\{v_k\}$; the photonic state is a classical mixture in which the $v_k$ mode is occupied with probability $\lambda_k$.
Thus, the overall fidelity and success probability can be calculated by replacing $\vab{f(\Delta)}^2$ with $\sum_k \lambda_k \vab{v_k(\Delta)}^2$, where $v_k(\Delta)$ is the Fourier transform of $v_k(t)$.
Finally, by using the following relation for an arbitrary function $h(\Delta)$
\begin{equation} \label{eq:g_relation}
    \begin{aligned}
        & \int \dd{\Delta} \sum_k \lambda_k \vab{v_k(\Delta)}^2 {h}(\Delta) \\
        &=  \frac{1}{2\pi}\iint \odif{t}\dd{t^\prime} g^{(1)}(t, t^\prime) \int \dd{\Delta} {h}(\Delta) e^{-i \Delta(t-t^\prime)},
    \end{aligned}
\end{equation}
we can evaluate the fidelity and success probability of the sequential-CAPS protocol directly from the autocorrelation function without the eigenmode decomposition.

\subsection{Practical photon source: single atom in an optical cavity}
As an exemplary photon source, we consider a cavity-QED-based source.
As illustrated in Fig.~\ref{fig2}(a), a simple three-level $\Lambda$-type atom is within an optical cavity, and a classical laser field drives the transition $\ket{u}_a \leftrightarrow \ket{e}_a$ with a time-dependent Rabi frequency $\Omega(t)$, which controls the excitation amplitude.
Simultaneously, the transition $\ket{e}_a \leftrightarrow \ket{g}_a$ is coupled to the cavity mode with a coupling strength $g$, enabling the emission of a photon into the cavity field that is then leaked out from the cavity at a rate $\kappa_\mathrm{ex}$.
This coherent combination of laser and cavity couplings enables the generation of single photons with well-defined temporal profiles, such as a Gaussian shape, at high probability~\cite{Vasilev2010, Utsugi2022}.

In this case, a major source of imperfection in the generated photon is the reexcitation, where the excited atom spontaneously decays back to $\ket{u}_a$ and is subsequently reexcited for photon emission with a different temporal profile than the desired one.
This results in mixed temporal modes with reduced purity~\cite{Maraner2020, Tanji2024, Kikura2025_high_purity}.
To quantitatively analyze the reexcitation effect, we numerically simulate the master equation that the source atom-cavity system follows to directly obtain $g^{(1)}(t,t^\prime)$ (see Appendix~\ref{s_sec:photon_generation} for the details of the calculation method).
We show exemplary results in Figs.~\ref{fig2}(b, c), to clearly illustrate the impact of the reexcitation process.
Here, we set the time-dependent Rabi frequency $\Omega(t)$ following the analytical expression in Ref.~\cite{Utsugi2022} that allows the generation of a photon with a Gaussian wavepacket.
The autocorrelation function should display a bivariate Gaussian function in the case of no reexcitation (the branching ratio $p_\text{br} = 0$).
However, finite $p_\mathrm{br}$ results in a small tail in the upper right due to the reexcitation effect that results in delayed photon excitation with a disturbed temporal mode.
The eigenmode decomposition of $g^{(1)}(t,t')$ in Eq.~\eqref{eq:g1-decomp} further reveals the fractional occupation of distinct temporal modes, as shown in Fig.~\ref{fig2}(c).

The photon-source information $g^{(1)}(t,t^\prime)$ is fed into the overall performance evaluation by using the relation~\eqref{eq:g_relation}, as shown in Figs.~\ref{fig2}(d, e).
Here, the control pulse $\Omega(t)$ is again shaped to generate a photon with a Gaussian temporal envelope, robust against temporal mode mismatch~\cite{Rohde2005}, with a width $\sigma_t$.
For the success probability, Fig.~\ref{fig2}(d) demonstrates the enhancement by replacing the photonic Hadamard gate in the well-known setup of the remote two-qubit gate with the $X$ gate for Bell-state generation.
This also shows that overall success probabilities saturate at $\sigma_t \gtrsim 1/\gamma$ to $\bar{P}_\text{gen} P_\text{CAPS}^\text{opt}$, with~\cite{Goto2019}
\begin{equation} \label{eq:P_gen^mux}
    \bar{P}_\text{gen} = \frac{\kappa_\text{ex}}{\kappa}\frac{g^2}{g^2 + (1-p_\text{br})\kappa\gamma}.
\end{equation}
For a larger branching ratio, the photon generation probability increases~\cite{Law1997,Goto2019} while the infidelity increases due to the reduced photon purity [see Fig.~\ref{fig2}(e)].
Even in the case of high $p_\mathrm{br}$, it is evident that a longer-pulse photon enables the infidelity to be below $10^{-4}$; we emphasize that this is evaluated with a realistic photon source having a single-photon trace purity~\cite{Fischer2018, Trivedi2020} of $M_\text{s} \coloneqq \sum_k \lambda_k^2/(\sum_k \lambda_k)^2 = 0.97$ (at $\sigma_t = 5/\gamma$ and $p_\text{br} = 0.4$)---demonstrating the inherent robustness of the CAPS-based protocol against an imperfect photon source.
This is in stark contrast to the two-photon-interference scheme, which requires near-unity single-photon purity, as the fidelity is determined as (see Appendix~\ref{ap_sec:HEG} for the derivation, as well as Ref.~\cite{Craddock2019, Krutyanskiy2023})
\begin{equation} \label{eq:F_ee}
    F_\text{ee} = \frac{1+M_s}{2}.
\end{equation}
In contrast, the CAPS framework maintains high fidelity even in the presence of realistic, non-ideal photon sources, allowing orders-of-magnitude improvement of the infidelity, for example, below 0.99 for two-photon interference and $10^{-4}$ level for CAPS protocol, for the case of $M_s = 0.97$ as shown above.

\section{Hardware-efficient hybrid emission-CAPS protocol}

\begin{figure}[t]
    \centering
    \includegraphics[width=0.99\linewidth]{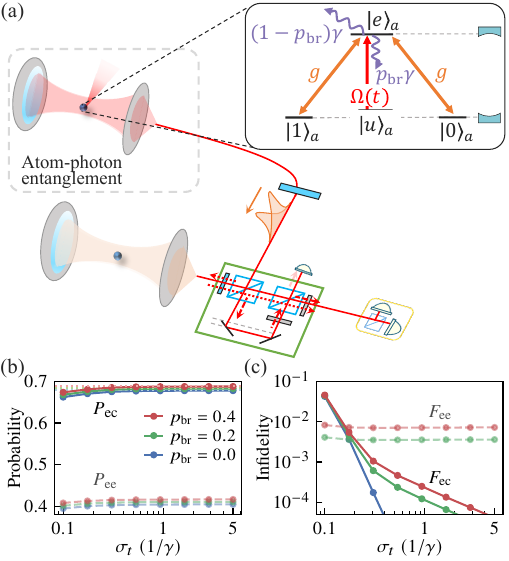}
    \caption{Performance of the hybrid emission-CAPS networking.
    (a) Schematic of the configuration consisting of the atom-photon entanglement generation and the memory loading.
    (b, c) Success probability and infidelity of the hybrid networking incorporating the imperfection of the initial atom-photon entanglement generation process.
    The system parameters are the same as Figs.~\ref{fig2}(d, e), leading to the upper bound $\bar{P}_\text{gen}^{\prime}$ of the atom-photon entanglement generation probability obtained from \eqref{eq:P_gen^mux} by replacing $(g,\kappa,\kappa_\text{ex})$ with $(2g,2\kappa,2\kappa_\text{ex})$, respectively.
    The dashed lines show the performance of photon-interference-based networking for reference, obtained using the same atom-cavity systems to generate the atom-photon entanglement.
    In panel (b), the dotted lines show the upper bound of the success probability $\bar{P}_\text{gen}^\prime P_\mathrm{CAPS}^\text{opt}$.
    }
    \label{fig3}
\end{figure}

From the results of the previous section, we propose a more hardware-efficient protocol for atom-atom entanglement generation, motivated by the inherent robustness of CAPS to finite photon impurity.
In this protocol, the first cavity is used to generate atom-photon entanglement, while the second cavity functions as a memory-loading interface for a photon entangled with the first atom, resulting in a heralded generation of remote atom-atom entanglement.
This eliminates the need for an independent photon source, as required in the protocol of the previous section, while achieving the same task of generating remote entangled atom pairs.

The proposed hybrid networking is illustrated in Fig.~\ref{fig3}(a).
For the atom in the first cavity, we consider a simple four-level system~\cite{Reiserer2015}.
A time-varying excitation laser is applied to the cavity-coupled atom for atom-photon entanglement generation, resulting in $\ket*{\Phi^{+}}_{ap} = (\ket{0}_a \ket*{0}_p + \ket{1}_a \ket*{1}_p)/\sqrt{2}$.
The photon is sent to Bob's cavity, and then qubit teleportation is executed using the CAPS gate, i.e., a CAPS gate on Bob's side followed by an $X$-basis measurement of the photon~\cite{Raymer2024} and a Hadamard gate applied to Bob's qubit, thereby yielding one of the atom–atom Bell states $\ket*{\Phi^\pm}_a = (\ket{0}_a\ket{0}_a \pm \ket{1}_a\ket{1}_a)/\sqrt{2}$ depending on the detection outcome (see Appendix~\ref{s_sec:memory_loading} for the details of the memory loading).

More precisely, for a photon with a frequency profile $f(\Delta)$, the click of the detector $j = 0(1)$ for the photonic state $\ket{+(-)}_p$ projects the two atoms onto
\begin{equation}
    \hat{\rho}_\text{ec}^{(j)} = \frac{1}{P_\text{ec}^{(j)}}\int \dd{\Delta} \frac{\vab{f(\Delta)}^2}{2} \hat{E}^\text{B}_a(\Delta)\ketbra*{\Phi_\text{id}^{(j)}}{\Phi_\text{id}^{(j)}} [\hat{E}^\text{B}_a(\Delta)]^\dagger,
\end{equation}
with the detection probability
\begin{equation} \label{eq:P_ec^(j)}
    P_\text{ec}^{(j)} = \int \dd{\Delta} \frac{\vab{f(\Delta)}^2}{2} \braket*[3]{\Phi_\text{id}^{(j)}}{[\hat{E}^\text{B}_a(\Delta)]^\dagger\hat{E}^\text{B}_a(\Delta)}{\Phi_\text{id}^{(j)}},
\end{equation}
where $\ket*{\Phi_\text{id}^{(0)}} = \ket*{\Phi^{-}}$, $\ket*{\Phi_\text{id}^{(1)}} = \ket*{\Phi^{+}}$.
Here, $\hat{E}^\text{B}_a(\Delta)$ represents the error operator acting on Bob's qubit induced by the frequency dependence of the reflection functions~\eqref{eq:freq-dependent-reflectance}, of which the explicit form is (see Appendix~\ref{ap_sec:HEG} for the details)
\begin{equation} \label{eq:error_op}
    \hat{E}_a^{\text{B}}(\Delta) = \frac{r_\text{m}^\text{B} + r^\text{opt,B}}{2}\hat{I}_a + \frac{r_\text{m}^\text{B} - r^\text{opt,B}}{2}\hat{Z}_a + \mathcal{O}(\Delta^2),
\end{equation}
such that the calibration of the mirror reflectivity as $r_\text{m}^\text{B} = r^\text{opt,B}$ results in the complete elimination of the first-order effect on the fidelity, $\hat{E}_a^{\text{B}}(\Delta) = r^\text{opt,B}\hat{I}_a + \mathcal{O}(\Delta^2)$.

Similarly to the previous section, this protocol also demonstrates robustness against photon impurity.
To see this, we characterize the emitted photon with the autocorrelation function~\eqref{eq:def_g1} for an end-to-end performance evaluation (see Appendix~\ref{s_sec:photon_generation} for the details).
In Figs.~\ref{fig3}(b, c), we show the end-to-end success probability $P_\text{ec} = \sum_{j=0,1}P_\text{ec}^{(j)}$ and average fidelity,
\begin{equation} \label{eq:F_ec}
    F_\text{ec} = \frac{P^{(0)}_\text{ec} \braket*[3]{\Phi^{-}}{\hat{\rho}^{(0)}_\text{ec}}{\Phi^{-}} + P^{(1)}_\text{ec} \braket*[3]{\Phi^{+}}{\hat{\rho}^{(1)}_\text{ec}}{\Phi^{+}}}{P^{(0)}_\text{ec}+ P^{(1)}_\text{ec} },
\end{equation}
which includes the effect of the source imperfection, together with the corresponding performance for two-photon-interference-based protocol~\cite{Beukers2024} where the two cavities are configured to create atom-photon entanglement via photon emission, before the photon pair is measured in Bell basis in the middle using beamsplitters and photon detectors (see Appendix~\ref{ap_sec:HEG} for the details).

Hybrid networking presents much higher success probabilities than photon-interference-based networking, mainly due to the fact that it avoids the 50\% upper bound of the two-photon-interference-based Bell measurement~\cite{Calsamiglia2001}; for an atom-photon entanglement generation probability $P_\text{gen}^{\prime}$, the success probability of the two-photon-interference protocol is
\begin{equation} \label{eq:P_ee}
    P_\text{ee} = \frac{(P_\text{gen}^{\prime})^2}{2},
\end{equation}
while that of the hybrid protocol is approximately $P_\text{gen}^{\prime}P_\text{CAPS}^\text{opt}$.
Moreover, Fig.~\ref{fig3}(c) shows that hybrid networking is highly robust against photon impurity---a feature also shared by sequential-CAPS-based networking---since neither protocol relies on two-photon interference.\footnote{An excitation pulse with a duration much shorter than the excited-state lifetime and large $\kappa_\text{ex}$ allows high-fidelity networking with photon interference, albeit with a much reduced photon emission probability~\cite{Li2024}.}

\section{Time- and wavelength-multiplexed CAPS networking}

\begin{figure}[t]
    \centering
    \includegraphics[width=0.99\linewidth]{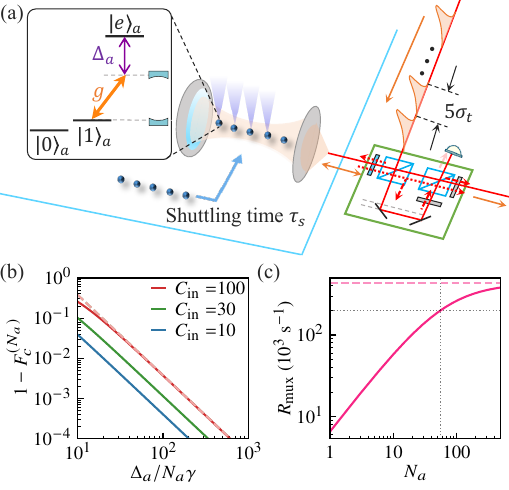}
    \caption{Multiplexed CAPS operation.
    (a) Schematic of the time-multiplexed operation.
    For an efficient use of the channel, a large number of atoms (atom number $N_a$) are shuttled to the cavity in parallel, followed by the application of hiding beams to all but one atom that performs a CAPS gate with an incoming photon with temporal width $\sigma_t$.
    After the time window $5\sigma_t$ for the first photon arrival, the hiding beam pattern is switched such that another atom can then interact with the next incoming photon.
    Once all atoms interact with their respective photon, the atom array is transported out while the new array is brought into the cavity mode for the next batch of operation.
    This operation is highly efficient for larger $N_a$, while the network rate saturates for $\tau_s \ll 5\sigma_t N_a$ [see panel (c)].
    (b) Crosstalk-induced infidelity of CAPS gates, evaluated by using Eq.~\eqref{app_eq:def_of_conditional_infid_of_N_a_atoms} for $N_a=200$ (solid lines), which agrees well with the approximate expression given by Eq.~\eqref{eq:infidelity_vs_detuning} (dashed lines).
    For high internal cooperativity $C_\mathrm{in} = 100$, choosing $\Delta_a/(N_a\gamma) > 2\times10^2$ keeps the crosstalk error below $10^{-3}$.
    (c) Remote entanglement generation rate~\eqref{eq:overall_mux_rate} with time-multiplexed hybrid emission-CAPS networking (Fig.~\ref{fig3}) for varying $N_a$.
    Here, we use realistic parameters summarized in Table~\ref{tab:parameter_summary_for_time_multiplexed_operation}, which motivated by $^{171}\mathrm{Yb}$ atoms coupled to a low-loss nanophotonic cavity.
    The dashed lines represent the intrinsic rate $R_\text{int}$.
    }
    \label{fig4}
\end{figure}

Even with the CAPS protocols presented above, which exhibit high performance and robustness to errors, the realistic implementation of quantum networks incurs additional time and operational constraints that substantially limit the overall network performance.
For example, qubit shuttling is required to bring initialized qubits to the cavity for trapped-atom platforms such as neutral atoms~\cite{Hartung2024} and trapped ions~\cite{Canteri2025}, which incur shuttling-time delays $\tau_s$ of up to hundreds of microseconds and limit the entanglement generation rates to $1/\tau_s$, orders of magnitude slower than the upper bound given by the success probability and the photon pulse duration~\cite{Li2024,Huie2021,Sunami2025}.
A practical solution is the time-multiplexed operations with a cavity that hosts a large number of individually addressable atoms~\cite{Hartung2024}, with near-term designs recently proposed for operating over $200$ atoms~\cite{Li2024,Sunami2025}.
A crucial requirement of the high-fidelity time-multiplexed operation is the careful management of the crosstalk effect, achievable by inducing ac Stark shifts to spectator qubits with addressable laser beams~\cite{Hu2025,Li2024}.
In this section, we first analyze the crosstalk error of CAPS gates in the presence of a large number of spectator atoms that are detuned from the cavity resonance, obtaining a simple analytical expression for the detuning required for time-multiplexed CAPS operations.
We then extend the analysis to the case of wavelength multiplexing, exploiting the multiple resonant frequencies naturally available, separated by the free spectral range.
These modes are well separated but still accessible by the large ac Stark shifts, thereby allowing low-crosstalk parallel operation of the CAPS-based networking with a single optical cavity.

\begin{table}[t]
    \caption{Summary of parameters for time-multiplexed CAPS networking.}
    \label{tab:parameter_summary_for_time_multiplexed_operation}
    \begin{ruledtabular}
        \small
        \begin{tabular}{lll}
            \multicolumn{3}{c}{Single-atom operation parameters} \\
            Parameter & Symbol & Value \\\hline
            Intrinsic finesse & $\mathcal{F}_\text{int}$ & 2000  \\
            Free spectral range & $\omega_\text{FSR}/2\pi$ & \qty{0.97}{\GHz} \\
            Internal cooperativity & $C_\mathrm{in}$ & 89 \\
            Cavity length & $L_\text{cav}$ & \qty{11}{\cm} [from Eq.~\eqref{eq:L_cav^opt}]\\
            Atomic decay rate & $\gamma/2\pi$ &  \qty{0.24}{\MHz} \\
            Internal cavity loss rate & $\kappa_\mathrm{in}/2\pi$ & \qty{0.24}{\MHz} [from Eq.~\eqref{eq:condition_for_kappa_in_gamma}] \\
            External coupling rate & $\kappa_\mathrm{ex}/2\pi$ & \qty{3.2}{\MHz} [from Eq.~\eqref{eq:kappa_ex_opt}] \\
            Atom-photon coupling rate & $g/2\pi$ & \qty{3.2}{\MHz} \\
            Photon pulse width & $\sigma_t$ & \qty{310}{\ns} \\
            Branching ratio & $p_\text{br}$ & 0.35 \\
            Impurity error & $1-F_\text{ec}$ & \num{5e-4} \\
            Success probability & $P_\text{ec}$ & 0.67 \\
            Intrinsic rate & $R_\text{int}$ & \qty{430}{\kHz} \\
        \end{tabular}
        \vspace{0.5em}
        \begin{tabular}{lll}
            \multicolumn{3}{c}{Time-multiplexed operation} \\
            Parameter & Symbol & Value \\\hline
            Shuttling time & $\tau_s$ & \qty{100}{\us} \\
            Number of atoms & $N_a$ & 56 \\
            Light shift & $\Delta_a/2\pi$ & \qty{3.9}{\GHz} \\
            Crosstalk error & $1-{F}_c^{(N_a)}$ & \num{4e-4} [from Eq.~\eqref{eq:infidelity_vs_detuning}] \\
            Total rate & $R_\text{mux}$ & \qty{200}{\kHz} [Fig.~\ref{fig4}(c)]
        \end{tabular}
    \end{ruledtabular}
\end{table}

\subsection{Time multiplexing}
In time-multiplexed operation, we prepare $N_a$ atoms in the cavity and operate a CAPS gate only on one of them, which we label a target atom index $k$, while the other $N_a-1$ atoms are shifted out of resonance by an amount $\Delta_a$, as shown in Fig.~\ref{fig4}(a).
This operation is repeated for each target atom $k$ ranging from 1 to $N_a$, allowing each of the atoms to try the CAPS gate once.
In this case, the reflection coefficients with $\Delta=0$ are
\begin{equation}\label{eq:reflection_coef_Naatom}
    r_j^{(m)} = 1- 2\kappa_\text{ex}\ab(\kappa + \frac{jg^2}{\gamma} + \frac{m g^2}{\gamma + i\Delta_a})^{-1} \quad (j \in \{0,1\}),
\end{equation}
where $m (\leq N_a-1)$ counts the number of spectator atoms being in state $\ket{1}_a$, and the last term $mg^2/(\gamma+i\Delta_a)$ represents the crosstalk effect due to the residual coupling between spectator atoms and the cavity mode.

To quantify crosstalk-induced infidelity, we model the CAPS gate as a quantum channel acting on the photonic qubit and a register of $N_a$ atoms (see Appendix~\ref{s_sec:tbm_crosstalk_effect} for the formal definition).
Here, we consider a sufficiently long pulse such that the absence of crosstalk results in the channel fidelity ${F}_c^{(N_a)} = 1$, i.e., the reflection amplitudes are calibrated to $r_\text{m} = r^\text{opt}$; the crosstalk-induced error is therefore quantified by the infidelity in the presence of the crosstalk effect, $1-{F}_c^{(N_a)}$.
For $\vab{\Delta_a} \gg N_a g^2/\kappa, \gamma$, the resulting infidelity approximates to (see Appendix~\ref{s_sec:tbm_crosstalk_effect} for the derivation)
\begin{equation} \label{eq:infidelity_vs_detuning}
    1- {F}_c^{(N_a)} \approx \frac{1}{2}\ab(1+\frac{3}{4}C_\text{in})\ab(\frac{N_a \gamma}{\Delta_a})^2,
\end{equation}
which is in excellent agreement with the exact result applicable for finite detuning, as shown in Fig.~\ref{fig4}(b).
Since Eq.~\eqref{eq:infidelity_vs_detuning} describes the fidelity of the operation performed on $N_a$ atoms and a photon, the average fidelity measure relevant for a single atom involved in one CAPS gate is approximately $\bigl[F_c^{(N_a)}\bigr]^{1/N_a}$.

In time-multiplexed operation, the $N_a$ qubits are prepared in the cavity, e.g., by qubit shuttling, before the CAPS gate is applied sequentially to each atom while switching the addressable hiding laser beams to the remaining $N_a-1$ atoms before being moved out for subsequent operations.
In this work, we consider the simplest case of time multiplexing where each qubit interacts with the channel only once, for a total of $N_a$ CAPS protocols for the case of a $N_a$-qubit register, while in-cavity qubit resets can further enhance the overall entanglement generation rates~\cite{Li2024,Sunami2025,Huie2021}.
In this case, the accumulated error per atom after the $N_a$ CAPS operations approximates to $1-{F}_c^{(N_a)}$.
Assuming a shuttling time $\tau_s$, atom number $N_a$, pulse separation of $5\sigma_t$,\footnote{This ensures the overlap of two successive photon temporal modes below $10^{-3}$.} and a success probability $P$ for remote entanglement protocols, the resulting network rate is
\begin{equation}\label{eq:overall_mux_rate}
    R_\text{mux}(N_a) = \frac{N_a P}{\tau_s + 5\sigma_t N_a} = R_\text{int}\frac{\tau_s/(5\sigma_t)}{\tau_s/(5\sigma_t) + N_a},
\end{equation}
while the intrinsic rate is $R_\text{int} = P/(5\sigma_t)$.

This enables the evaluation of the required $N_a$ to achieve a desired network rates; then, the necessary detuning $\Delta_a$ is identified.
For a concrete evaluation, in Fig.~\ref{fig4}(c), we show the estimated rate of the hybrid emission-CAPS protocol with the parameters in Table~\ref{tab:parameter_summary_for_time_multiplexed_operation}, motivated by using the telecom ($^{3}\text{P}_0$--$^{3}\text{D}_1$) transition of $^{171}\text{Yb}$ atoms coupled to low-loss nanophotonic cavity~\cite{Horikawa2025,Sunami2025}.
Since we do not implement in-cavity qubit cooling and resets, no feedback operations on atoms within the cavity are required; the dominant time cost on the networking rate arises from the atom shuttling time $\tau_s$; the case of the single-atom operation yields a rate of $R_\text{mux}(1) = \qty{6.6}{\kHz}$, a factor of 65 smaller than the intrinsic rate of $R_\text{int} = \qty{430}{\kHz}$.
In contrast, using $N_a = 56$ atoms efficiently enhances the total rate to \qty{200}{\kHz}.
\begin{figure*}[t]
    \centering
    \includegraphics[width=0.95\linewidth]{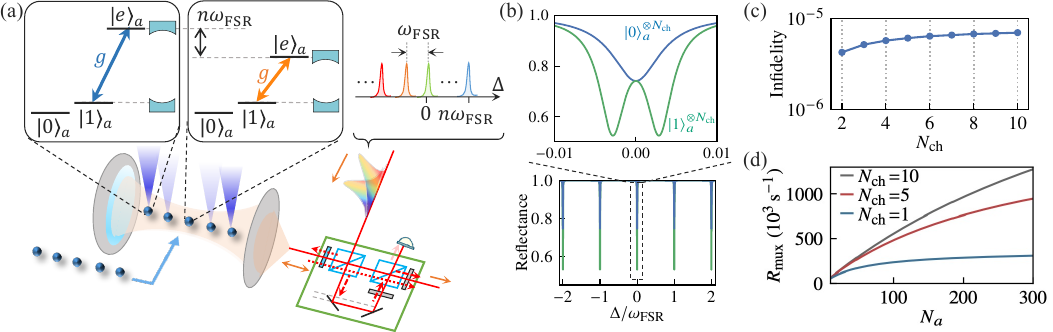}
    \caption{Wavelength-multiplexed CAPS operations.
    (a) Schematic of wavelength-multiplexed cavity-QED systems where each atom in $\ket{1}_a$ couples to the different cavity modes by tuning the resonant frequency of the atoms via ac Stark shift.
    (b) Cavity reflection spectra evaluated by the transfer matrix method.
    $N_a = 10$ atoms are assigned to $N_\text{ch} = 10$ channels respectively, and we plot the reflectance for the cases with all the atoms prepared in $\ket{0}_a$ (blue line) and $\ket{1}_a$ (green line).
    The plot on top provides a magnified view of one representative mode.
    (c) Crosstalk effect for the wavelength-multiplexed CAPS gates (see Appendix~\ref{s_sec:transfer-matrix_approach} for the details).
    Here, we consider the same setup in Fig.~\ref{fig4}(c) for the time-multiplexed operation, where the key parameters are summarized in Table~\ref{tab:parameter_summary_for_time_multiplexed_operation}.
    (d) Time-multiplexed entanglement generation rates with multiple wavelength channels. The $N_a$ atoms are partitioned into $N_\mathrm{ch}$ channels for parallel execution of time-multiplexed entanglement generation for each channel.
    }
    \label{fig5}
\end{figure*}
In this case, the required light shift to realize the crosstalk error of \num{4e-4} is $\Delta_a/2\pi = \qty{3.9}{\GHz}$ from Eq.~\eqref{eq:infidelity_vs_detuning}.
Since the spectator atoms remain in the ground state for the ideal operation, shining a hiding beam on the transition between $\ket{e}_a$ and the other excited state enables a large light shift with favorable scattering error scaling, as demonstrated for qubit measurements~\cite{Hu2025, Bluvstein2026} and qubit gates~\cite{Burgers2022}.
More concretely, for $^{171}\text{Yb}$ atoms, by using the $^{3}\text{D}_1$ ($\ket{e}_a$) to $6s8p\,^{3}\text{P}_1$ transition at $\lambda_\mathrm{hiding} = $ \qty{522}{\nm},
the hiding-beam detuning of around \qty{1}{\GHz} suppresses the scattering error to $\approx 10^{-4}$ with negligible heating and decoherence of the atoms, as discussed in Ref.~\cite{Li2025} for remote entanglement generation via time-multiplexed atom-photon entanglement emissions; a moderate laser power of around \qty{180}{\uW} with a $1/e^2$ beam radius of around \qty{500}{\nm} is required for the light shift of $\Delta_a$.
In addition, the mm-scale interface region for atom-photon coupling over 11-cm cavity length has enough space to arrange $N_a=56$ atoms with a sufficiently large spacing to ensure negligible crosstalk on neighboring atoms~\cite{Hu2025}.
Overall, remote entanglement generation at a rate of \qty{200}{\kHz} with infidelity of \num{e-3}, which includes errors arising from photon impurity, crosstalk between multiple atoms, and hiding beams, is achievable with the realistic hardware setup (see Table~\ref{tab:parameter_summary_for_time_multiplexed_operation}).
We finally mention that increasing the number of atoms further enhances the total rate, as shown in Fig.~\ref{fig4}(c), at the cost of a larger light shift required.

\subsection{Wavelength multiplexing}
While time multiplexing offers substantially faster remote entanglement generation than native implementation with a single-qubit network register, the rate is inherently limited by the fact that the atom-photon interaction must operate sequentially.
This is because we have so far only considered a specific cavity resonance and neglected others, which are typically far off-resonant from the atomic transition.

We now utilize the multiple resonant frequencies available in optical cavities and treat them as separate wavelength channels, further enhancing the single-cavity network performance.
In particular, we consider the cases where the cavity is sufficiently long and the free-spectral range $\omega_\mathrm{FSR}$ is relatively small, such that atomic resonances can be shifted between different resonant modes by laser beams.
It is also crucial that the finesse is sufficiently high, ensuring that each mode is spectrally well isolated; such an operating regime is realized with several optical cavity implementations, such as Fabry-P\`{e}rot cavities~\cite{Aqua2026} and nanofiber cavities~\cite{Horikawa2025}.

For such a wavelength-multiplexed parallel networking to be realistic, it is essential that the errors arising from cross-channel crosstalk be sufficiently low.
More concretely, we evaluate the effect of resonantly coupled atoms in an adjacent mode on the response of the target atom-cavity system.
While a full multi-atom, multi-mode simulation of the coupled atom-cavity system is computationally intractable, here we employ the transfer-matrix approach~\cite{Nemet2020} where we linearize the atomic response inside the cavity and treat the entire atom-cavity system as a sequence of input-output elements expressed by $2\times2$ matrices (see the details in Appendix~\ref{s_sec:transfer-matrix_approach}).
This precisely evaluates the reflection function, provided the input light consists of up to one photon~\cite{Wang2025}, which is compatible with the quantitative evaluation of the CAPS protocol.

As illustrated in Figs.~\ref{fig5}(a, b), we are interested in the reflection coefficients of atom-cavity systems with $N_\mathrm{ch}$ atoms, each shifted by a different amount to be coupled to distinct resonance modes of the cavity at frequency $(n_0 + n)\omega_\text{FSR}~(n \in \mathbb{Z})$, where $n_0$ is the central mode index.
We then evaluate the average infidelity of the CAPS operation in the presence of $N_\mathrm{ch}-1$ spectator atoms coupled to nearby resonance modes---similar to the analysis in time-multiplexed operation---as shown in Fig.~\ref{fig5}(c) for an exemplary case of $\omega_\mathrm{FSR}/2\pi = \qty{0.97}{\GHz}$ and intrinsic finesse $\mathcal{F}_\mathrm{int} = 2000$.
We chose this as an accessible regime both in terms of light shift capability and cavity parameters, where we find that the cross-channel crosstalk effect is negligible below \num{e-5}; this suggests that each mode can be treated individually, allowing parallel networking.
We note that even for the moderate intrinsic finesse of $500$, the average cross-channel crosstalk infidelity remains below $10^{-4}$ (see Appendix~\ref{s_sec:transfer-matrix_approach}), making this approach an attractive option for a wide variety of optical cavity designs.

\subsection{Overall performance}
To fully utilize both time- and wavelength-multiplexed operations, we evaluate zoned multiplexing~\cite{Sunami2025} where the time-multiplexed operation is performed over multiple independent operating sets of qubits for each wavelength channel.
We set the total atom number in the cavity to be $N_a$, which is partitioned into $N_\mathrm{ch}$ optical channels available for parallel entanglement generation trials.
We then consider the parallel execution of the time-multiplexed operation with $\lfloor N_a/N_\mathrm{ch} \rfloor$ atoms at a rate of $R_\mathrm{mux}(\lfloor N_a/N_\mathrm{ch} \rfloor)$, obtaining the total network rate $N_\mathrm{ch} R_\mathrm{mux}(\lfloor N_a/N_\mathrm{ch} \rfloor)$.
The achievable entanglement generation rate is plotted in Fig.~\ref{fig5}(d), showing a rapid increase in the overall rate $R_\mathrm{mux}$ for increased $N_\mathrm{ch}$, approaching $\qty{e6}{\second^{-1}}$ with a few hundred atoms and 10 channels, nearly a factor of 5 increase over single-channel operation without additional cavities or atoms.
A successful integration of this approach significantly improves the network performance of a \emph{single} optical cavity; thus, it is an attractive alternative to physical channel multiplexing requiring multiple optical cavities to scale the network rates~\cite{Sunami2025,Sinclair2025}.

\section{Conclusion and Outlook}

In conclusion, we have established CAPS-based atom-photon gate operation as a promising primitive for high-rate, high-fidelity quantum networking and demonstrated its robustness to experimental imperfections.
The key to this advancement is the careful incorporation of error cancellation methods supported by thorough modeling of the optical response of atom-cavity systems, including the crosstalk effects in 200-atom systems for time multiplexing.
As an example, for the case of the telecom-band transition of ${}^{171}$Yb atoms~\cite{Li2024, Sunami2025, Li2025}, we estimate a rate of \qty{2e5}{s^{-1}} at a heralded fidelity of 0.999, just with a single round of entanglement generation trials for each atom.
This is in contrast to the time-multiplexed operations considered previously, with many rounds of trials necessary to reach high rates through repeated qubit reinitialization and cooling~\cite{Li2024,Huie2021,Sunami2025}; the CAPS-based protocol achieves a higher rate with only light shift laser beams needed for remote entanglement generation, without requiring high-power excitation lasers and other complex qubit controls---therefore, we predict this to be a viable option for cavity-based quantum interconnect.
We have further demonstrated that wavelength multiplexing, using multiple modes naturally accessible for optical cavities, scales the network performance without additional in-module hardware complexities.

We conclude with a few remarks on the further improvements of the CAPS operations, implications for the design of networked fault-tolerant quantum computers, and an application for long-distance quantum communication.

A major performance improvement of the CAPS gate is expected with the use of techniques already proposed or utilized for the two-photon-interference schemes.
An example is the use of photon detection time information: when the photon is detected at the end of the protocol, the timing information provides rich insight into the error characteristics of the generated atom-atom entanglement. For the case of photon-interference-based networking, detection time information provides error probabilities and error biases of generated Bell pairs~\cite{Li2024}, as well as a way for significant error suppression through detection time filtering~\cite{Kikura2025}.
This is also expected to be efficient for the CAPS protocol, as the infidelity sources of CAPS studied in this work are also time-dependent; this may become a crucial ingredient for achieving even better performance than that already analyzed in this work.

The improved performance of CAPS-based quantum networking over the two-photon-interference protocol, without needing fast, high-power excitation lasers, may transform the architectural design of multiprocessor fault-tolerant quantum computers~\cite{Sunami2025}.
With only atom shuttling and light shift beams required for passive interconnect operation, and given the efficiency of entanglement distillation~\cite{Pattison2024, Sunami2025, sunami2025boosting}, greater flexibility in module layout is expected.
Furthermore, the high success probability now allows the use of only a single round of entanglement generation trials while maintaining a good networking rate, thus eliminating the complicated conditional sequencing required to reset only the atoms that failed in the previous round~\cite{Li2024, Sunami2025}.
The full system design, involving the logical entanglement generation~\cite{Sunami2025transversal, Sinclair2025}, will thus be more efficient thanks to the simplicity and performance of the CAPS-based remote entanglement generation.

Finally, multiplexed CAPS-based memory loading, discussed in Appendix~\ref{s_sec:memory_loading}, is also a powerful scheme for long-distance quantum communication, including quantum repeater operation, thanks to the robustness of the CAPS gate to source and channel fluctuations, improved success probability, and high fidelity
(see Appendix~\ref{s_sec:optimize_caps_gate}).
For example, a variant of CAPS-based networking, with a single-photon source replaced by an entangled photon-pair source (see also Appendix~\ref{ap_sec:HEG}), offers advantages in extreme-loss communication settings, including the satellite-to-ground downlink assisted quantum networking~\cite{Ji2025}.

\subsection*{Data availability}
The data supporting the findings in this work are available upon reasonable request.

\begin{acknowledgements}
We thank J. Ji and C. Simon for extensive discussions on the quantum repeater implementation based on CAPS gates, and O. Rubies-Bigorda for contributions to the early stages of this work.
We acknowledge C. Simon, R. Inoue, and in particular, K. Nicolas Komagata for careful reading of the manuscript.
\end{acknowledgements}

\appendix

\section*{Appendices}
The appendices are organized as follows.
In Appendix~\ref{s_sec:optimize_caps_gate}, we define the conditional gate fidelity and success probability for CAPS gates and provide analytical expressions in both long-pulse and finite-bandwidth regimes.
Based on the analytical formula, in Appendix~\ref{s_sec:memory_loading} we analyze the performance of CAPS-based memory loading via photonic-state teleportation.
In Appendix~\ref{s_sec:photon_generation}, we formalize performance metrics for single-photon generation and its application to atom-photon entanglement generation, followed by Appendix~\ref{ap_sec:HEG} where we analyze heralded remote entanglement generation using CAPS gates, as well as photon-interference-based networking for comparison.
In Appendix~\ref{s_sec:tbm_crosstalk_effect}, we evaluate crosstalk in multi-atom CAPS operations and derive its scaling with the number of atoms and detuning.
Appendix~\ref{s_sec:transfer-matrix_approach} is for a transfer-matrix model for wavelength-multiplexed CAPS gates and evaluates channel crosstalk under realistic conditions.

\section{Optimization of CAPS gates} \label{s_sec:optimize_caps_gate}
We introduce two metrics for the CAPS gate, conditional gate fidelity and success probability, and use them to quantify the performance of the CAPS gate.
Although the CAPS gate suffers optical loss from atomic spontaneous emission and intracavity loss, many applications considered in this work, including photon-mediated remote atomic-qubit gates~\cite{Duan2005, Lin2006}, memory-loading schemes~\cite{Kalb2015, Raymer2024}, and CAPS-based remote entanglement generation, allow for the postselection of cases where photons were measured at the end of the protocol.
In such a heralded protocol, the conditional gate fidelity and the success probability are the relevant figures of merit for the CAPS gate.

In the following, we first introduce the general forms of conditional fidelity $F_c$ and success probability $P$ in Appendix~\ref{sec:general-cond-measure}.
Then, we evaluate $F_c$ and $P$ of the CAPS gate in the long-pulse limit in Appendix~\ref{s_subsec:caps_long_pulse_limit}.
Finally, we extend to the frequency-dependent behavior relevant to the high-speed operation of CAPS gates in Appendix~\ref{app:frequency-dep}.

\subsection{General framework for metrics to evaluate CAPS gates}\label{sec:general-cond-measure}
Let us consider the joint Hilbert space $\mathcal{H}^{ap} = \mathcal{H}^a \otimes \mathcal{H}^p$: the atomic subspace $\mathcal{H}^a$ is spanned by the orthonormal basis $\{\ket{0}_a, \ket{1}_a, \ket{e}_a, \ket{\tilde{o}}_a\}$,
while the photonic subspace $\mathcal{H}^p$ is spanned by
$\{\ket{0}_p, \ket{1}_p, \ket{\varnothing}_p\}$, where $\ket{\tilde{o}}_a$ represents an auxiliary state that can be populated via atomic decay from $\ket{e}_a$ in addition to the qubit states $\ket{0}_a$ and $\ket{1}_a$, and $\ket{\varnothing}_p$ denotes the vacuum state.
Following the standard leakage framework where the system of interest is embedded in a larger Hilbert space that also contains all loss pathways, we partition the atom–photon space $\mathcal{H}^{ap}$ into the direct sum, $\mathcal{H}^{ap} \cong \mathcal{X}_{\mathrm{q}} \oplus \mathcal{X}_{\mathrm{loss}}$, where
\begin{equation}
    \mathcal{X}_{\mathrm{q}} = \operatorname{span}\{\ket{0}_{a},\ket{1}_{a}\} \otimes \operatorname{span}\{\ket{0}_{p},\ket{1}_{p}\}
\end{equation}
represents the $d_\text{q}$-dimensional \textit{computational subspace}, whereas $\mathcal{X}_{\mathrm{loss}}$ (dimension $d_{\mathrm{loss}}$) is the \textit{loss subspace}, occupied when the photon leaks out.
The leakage $L$ of a channel $\mathcal{G}$ is defined by~\cite{Wood2018}
\begin{equation}
    \begin{aligned}
        L(\mathcal{G}) =& 1 - \int\dd{\psi_\text{q}} \Tr[\bm{1}_\text{q} \mathcal{G}(\ketbra*{\psi_\text{q}}{\psi_\text{q}})] \\
        =& 1- \Tr\ab[\bm{1}_\text{q}\mathcal{G}\ab(\frac{\bm{1}_q}{d_\text{q}})],
    \end{aligned}
\end{equation}
where the integral is taken over the Haar measure of all states $\ket*{\psi_\text{q}}$ in the computational subspace $\mathcal{X}_{\mathrm{q}}$ and $\bm{1}_\text{q}$ denotes the projector onto $\mathcal{X}_\text{q}$.
We define the average gate fidelity ${F}$ in the subspace $\mathcal{X}_\text{q}$ as
\begin{equation}
    {F}(\mathcal{G}, U_\text{tar}) = \int\dd{\psi_\text{q}} \braket*[3]{\psi_\text{q}}{\hat{U}_\text{tar}^\dagger\bm{1}_\text{q}\mathcal{G}(\ketbra*{\psi_\text{q}}{\psi_\text{q}})\bm{1}_\text{q}\hat{U}_\text{tar}}{\psi_\text{q}},
\end{equation}
where $\hat{U}_\text{tar}$ is the target unitary operator.
For the Kraus representation $\mathcal{G}(\hat{\rho}) = \sum_k \hat{G}_k\hat{\rho}\hat{G}_k^\dagger$, this reduces to~\cite{Pedersen2007}
\begin{equation} \label{app_eq:average_gate_fidelity}
    \begin{aligned}
        {F}(\mathcal{G}, U_\text{tar}) =& \frac{\sum_k \ab(\Tr[\bm{1}_\text{q}\hat{G}_\text{k}^\dagger\bm{1}_\text{q}\hat{G}_\text{k} \bm{1}_\text{q}] + |\Tr[\hat{U}_\text{tar}^\dagger\bm{1}_\text{q}\hat{G}_\text{k} \bm{1}_\text{q}]|^2)}{d_\text{q}(d_\text{q}+1)}, \\
        =& \frac{d_\text{q} F_\text{pro}(\mathcal{G}, U_\text{tar}) + 1- L(\mathcal{G})}{d_\text{q} + 1},
    \end{aligned}
\end{equation}
where we have used the process fidelity in the computational subspace,
\begin{equation}
    F_\text{pro}(\mathcal{G}, U_\text{tar}) = \frac{|\Tr[\hat{U}_\text{tar}^\dagger\bm{1}_\text{q}\hat{G}_\text{k} \bm{1}_\text{q}]|^2}{d_\text{q}^2}.
\end{equation}
When we postselect events where the gate output remains in the qubit subspace, the average success probability $P$ and the corresponding average conditional fidelity $F_c$ are given by~\cite{Pedersen2007}.
\begin{equation}\label{S_eq:leakage-fidelity}
    P=1-L, ~F_c=\frac{F}{1-L}.
\end{equation}
Using Eq.~\eqref{app_eq:average_gate_fidelity}, we find
\begin{equation} \label{app_eq:conditional_infidelity_with_F_pro_L}
    1-F_c = \frac{d_\text{q}}{d_\text{q}+1} \ab(1-\frac{F_\text{pro}}{1-L}).
\end{equation}

\subsection{Evaluation and optimization of CAPS gates in the long-pulse regime} \label{s_subsec:caps_long_pulse_limit}
For the CAPS gate, the target unitary operator is given by
\begin{equation} \label{S_eq:simplest_U_tar}
    \hat{U}_\text{tar} = \bm{1}_a \otimes \ketbra{0}[_p]{0} + (-\ketbra{0}[_a]{0}+\ketbra{1}[_a]{1}) \otimes \ketbra{1}[_p]{1},
\end{equation}
which corresponds to the CZ gate up to local Pauli gates.
We first consider the standard CAPS gate where the mirror perfectly reflects the photon.
For a sufficiently long photon pulse that has a narrow bandwidth, the reflection amplitudes in Eq.~\eqref{eq:freq-dependent-reflectance} can be approximated by their resonant ones, $r_i(0) \, (i\in \{0,1\})$.
The Kraus operator $\hat{G}_0$ that corresponds to the event without photon loss is given by
\begin{equation}
    \hat{G}_0 = \bm{1}_a \otimes \ketbra{0}[_p]{0} + (r_0 \ketbra{0}[_a]{0} + r_1 \ketbra{1}[_a]{1}) \otimes \ketbra{1}[_p]{1},
\end{equation}
where $r_0$ and $r_1$ are given in Eq.~\eqref{eq:freq-dependent-reflectance} with $\Delta = 0$.
All the other events project the photonic qubit onto the vacuum state.
Thus, the success probability and the conditional gate infidelity are given by
\begin{equation}
    \begin{gathered}
        P_\text{CAPS} = \frac{2 + |r_0|^2 + |r_1|^2}{4}, \\
        1-F_c = \frac{4}{5}\ab(1-\frac{|2-r_0+r_1|^2}{16P_\text{CAPS}}).
    \end{gathered}
\end{equation}
The optimization of the external coupling rate via Eq.~\eqref{eq:kappa_ex_opt} sets the reflectivities to $-r_0 = r_1 = r^\text{opt}$, which gives
\begin{gather}
    1- {F}_c = \frac{2}{5}\frac{1}{1+C_\text{in}}, \label{eq:standard_CAPS_F_c} \\
    {P}_\text{CAPS} = 1-\frac{\sqrt{1+2C_\text{in}}}{1+C_\text{in} + \sqrt{1+2C_\text{in}}}.
    \label{app_eq:standard_CAPS_P}
\end{gather}
This is the conventional performance of the CAPS gate widely studied, where infidelity of $<10^{-3}$ requires $C_\text{in} > 400$, which is beyond state-of-the-art optical cavity implementations.

To eliminate the reflectivity mismatch between two polarization modes, we deliberately introduce a calibrated loss in the $H$-polarized path, similarly to the idea of Ref.~\cite{Chen2021}, by turning the HWP away from $\theta_r = \pi/4$:
specifically, we set $\theta_r$ such that the reflection amplitude at the second PBS is $r_\text{m}$.
The corresponding Kraus operator becomes
\begin{equation}
    \hat{G}_0 = r_\text{m}\bm{1}_a \otimes \ketbra{0}[_p]{0} + (r_0 \ketbra{0}[_a]{0} + r_1 \ketbra{1}[_a]{1}) \otimes \ketbra{1}[_p]{1}.
\end{equation}
Then, the two measures are replaced with
\begin{equation} \label{eq:standard_CAPS_F_c_r_m}
    \begin{gathered}
        P_\text{CAPS} = \frac{2|r_\text{m}|^2 + |r_0|^2 + |r_1|^2}{4}, \\
        1-F_c = \frac{4}{5}\ab(1-\frac{|2r_\text{m}-r_0+r_1|^2}{16P_\text{CAPS}}).
    \end{gathered}
\end{equation}

\begin{figure}[t]
    \centering
    \includegraphics[width=0.65\linewidth]{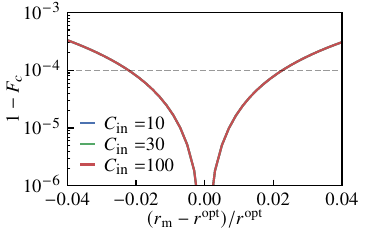}
    \caption{Sensitivity of the CAPS gate to the imperfect calibration of the mirror amplitude $r_\text{m}$. The infidelity $1-F_c$ in Eq.~\eqref{eq:standard_CAPS_F_c_r_m} is plotted as a function of the relative deviation $(r_\text{m}-r_\text{m}^\text{opt})/r_\text{m}^\text{opt}$ for different $C_\text{in}$.
    The results for $C_\text{in} = 10, 30, 100$ are degenerate in the plot.}
    \label{fig6}
\end{figure}

This shows that setting $r_\text{m} = -r_0 = r_1= r^\text{opt}$ results in $F_c^\text{opt} = 1$ with a finite reduction in success probability as $P_\text{CAPS}^\text{opt} = (r^\text{opt})^2 = 2P_\text{CAPS} -1$; the infidelity as a function of the relative deviation $(r_\text{m}-r^\text{opt})/r^\text{opt}$ is plotted in Fig.~\ref{fig6}.
This shows that the calibration of the mirror amplitude to ensure the relative deviation up to $2\%$ is sufficient to suppress the error below $10^{-4}$.
Crucially, this added loss can be heralded: a detector placed at the unused output port of the PBS, illustrated as a photodetector with a label ``Erasure detection'' in Fig.~\ref{fig1}(b), registers any $H$-polarized photon diverted for attenuation, thereby converting to an erasure of the photonic qubit.
The detector click at this port indicates that the photon did not interact with the cavity, and as such, the protocol can be retried immediately without time-consuming atom reinitialization.

\subsection{Frequency-dependent CAPS gate analysis} \label{app:frequency-dep}
The discussion in Appendix~\ref{s_subsec:caps_long_pulse_limit} relied on the long-pulse limit.
However, for fast networking, it is necessary to operate with short photonic pulses featuring relatively large bandwidths where the frequency-dependent response of the atom-cavity system must be taken into account.
The reflection functions of the cavity are~\cite{Cohen2018, Raymer2024, Utsugi2025}:
\begin{equation}
    \begin{aligned}
        {r}_0(\Delta) =& \frac{-\kappa_\text{ex} + \kappa_\text{in} -i\Delta}{\kappa_\text{ex}+\kappa_\text{in} - i\Delta}, \\
        {r}_1(\Delta) =& \frac{(-\kappa_\text{ex} + \kappa_\text{in} -i\Delta)(\gamma+i\Delta_a-i\Delta) + g^2}{(\kappa_\text{ex}+\kappa_\text{in} - i\Delta)(\gamma+i\Delta_a-i\Delta) + g^2},
    \end{aligned}\label{app_eq:freq-dep-reflection}
\end{equation}
where $\Delta = \omega-\omega_c$ is the detuning from the cavity frequency $\omega_c$ and $\Delta_a = \omega_a - \omega_c$ is the detuning of the atomic transition [We set $\Delta_a = 0$ in Eq.~\eqref{eq:freq-dependent-reflectance} of the main text for simplicity].
To incorporate the spectrum of the photon, we define a photonic qubit state with a spectral amplitude ${f}(\Delta)$ as
\begin{equation}
    \ket*{j; {f}}_p = \int \dd{\Delta} {f}(\Delta)\hat{a}_j^\dagger(\Delta)\ket{\varnothing}_p\quad (j\in \{0,1\}),
\end{equation}
where $\hat{a}_j(\Delta)$ is the annihilation operator of a monochromatic photon in the polarization mode $j$.
The inner product is given as
\begin{equation}
    _{p}\braket*{k; {h}}{j; {f}}_{p} = \delta_{kj}\aab*{{h}, {f}},
\end{equation}
with the inner product of functions,
\begin{equation}
    \aab*{{h}, {f}} = \int \dd{\Delta} {h}^\ast(\Delta) {f}(\Delta).
\end{equation}
We allow the norm of $\ket*{j; {f}}_p$ to be less than $1$:
\begin{equation}
    _p\braket*{j; {f}}{j; {f}}_p = \int \dd\Delta |{f}(\Delta)|^2 \leq 1,
\end{equation}
for the simplicity of notation in the following analysis.
The target unitary operator for the photon with a frequency mode $f$ is then given by replacing $\ket{j}_p$ with $\ket*{j;{f}}_p$ in Eq.~\eqref{S_eq:simplest_U_tar}, yielding
\begin{equation} \label{S_eq:mode_incorporated_target_U}
    \begin{aligned}
        \hat{U}_{\text{tar},f} =& \bm{1}_a \otimes \ketbra*{0;{f}}[_p]{0;{f}} \\
        &+ (-\ketbra{0}[_a]{0}+\ketbra{1}[_a]{1}) \otimes \ketbra*{1;{f}}[_p]{1;{f}}.
    \end{aligned}
\end{equation}
The corresponding Kraus operator $\hat{G}_0$ is given by
\begin{equation} \label{S_eq:mode_incorporated_G_0}
    \begin{aligned}
        \hat{G}_{0,f} =& r_{\text{m}}\bm{1}_a \otimes \ketbra*{0;{f}}[_p]{0;{f}}  \\
        &+ \sum_{j = 0,1} \ketbra{j}[_a]{j} \otimes \ketbra*{1; {f}_j}[_p]{1;{f}},
    \end{aligned}
\end{equation}
where we define
\begin{equation}
    {f}_j(\Delta) =  e^{-i\tau_{\text{m}} \Delta} {r}_j(\Delta) {f}(\Delta),
\end{equation}
and $e^{-i\tau_{\text{m}}\Delta}$ denotes the action of the delay line [see Fig.~\ref{fig1}(b)]; while the delay line is inserted in the path for the $\ket{0}_p$ photon, the effect is incorporated to the model here for the $\ket{1}_p$ photon by shifting the temporal origin, for notation simplicity.
By using Eqs.~\eqref{S_eq:mode_incorporated_target_U} and \eqref{S_eq:mode_incorporated_G_0}, we calculate
\begin{equation}
    \begin{aligned}
        \Tr[\hat{G}_{0,f}^\dagger \hat{G}_{0,f}] =& 2|r_{\text{m}}|^2 + \aab*{{f}_0, {f}_0} + \aab*{{f}_1, {f}_1}, \\
        \Tr[\hat{U}_{\text{tar},f}^\dagger \hat{G}_{0,f}] =& 2r_{\text{m}} - \aab*{{f}, {f}_0} + \aab*{{f}, {f}_1},
    \end{aligned}
\end{equation}
resulting in the process fidelity and the leakage as
\begin{equation} \label{S_eq:F_pro-L_f}
    \begin{aligned}
        F_{\text{pro},f} =& \frac{|2r_{\text{m}} - \aab*{{f}, {f}_0} + \aab*{{f}, {f}_1}|^2}{16}, \\
        L_f =& 1- \frac{2|r_{\text{m}}|^2 + \aab*{{f}_0, {f}_0} + \aab*{{f}_1, {f}_1}}{4},
    \end{aligned}
\end{equation}
which enables the evaluation of the conditional infidelity and the success probability by using Eqs.~\eqref{S_eq:leakage-fidelity} and \eqref{app_eq:conditional_infidelity_with_F_pro_L}.

\subsection{Mitigating pulse delay via cavity optimization} \label{app:pulse-delay}
Here, we outline one of the main sources of infidelity in the CAPS gate, temporal-mode mismatch caused by the atomic-state-dependent pulse delay, and discuss practical mitigation measures.
First, we derive explicit expressions for the atomic-state-dependent pulse delays in Appendix~\ref{app:pulse-delay-derivation}.
Second, we present a concrete example that employs a Gaussian waveform in Appendix~\ref{app:gaussian-model}.
Finally, we describe a practical method to mitigate temporal-mode mismatch by optimizing the cavity length in Appendix~\ref{app:cavity-length-optimization}.

\subsubsection{State-dependent pulse delay}\label{app:pulse-delay-derivation}
We consider that the atom is resonantly coupled to the cavity, $\Delta_a=0$, and perform the Taylor expansion of reflection functions of Eq.~\eqref{app_eq:freq-dep-reflection} as
\begin{equation} \label{S_eq:taylor_expansion_of_refelection_functions}
    \begin{aligned}
        {r}_0(\Delta) =& r_0 - i \frac{2\kappa_\text{ex}}{\kappa^2}\Delta + \frac{2\kappa_\text{ex}}{\kappa^3}\Delta^2 + \mathcal{O}(\Delta^3), \\
        {r}_1(\Delta) =& r_1 + i\frac{2\kappa_\text{ex}(g^2-\gamma^2)}{(g^2+\kappa\gamma)^2}\Delta \\
        &- \frac{2\kappa_\text{ex}(g^2\kappa + 2g^2\gamma - \gamma^3)}{(g^2+\kappa\gamma)^3}\Delta^2 + \mathcal{O}(\Delta^3).
    \end{aligned}
\end{equation}
For a sufficiently small $\Delta$ such that we can neglect the second- and higher-order terms, we find
\begin{equation} \label{S_eq:r_incorporating_pulse_delay}
    {r}_j(\Delta) =  {r}_j(0) + {r}_j^\prime(0)\Delta + \mathcal{O}(\Delta^2) = r_j e^{i\tau_j \Delta} + \mathcal{O}(\Delta^2),
\end{equation}
where $\tau_j = -i {r}_j^\prime(0)/{r}_j(0)$ represents the pulse delay induced by the reflection off the cavity~\cite{Tomm2023, Utsugi2025}, as shown in Fig.~\ref{fig7}(a).
The explicit forms are given by
\begin{equation}
    \begin{aligned}
        \tau_0 =& \frac{2\kappa_\text{ex}}{\kappa_\text{ex}^2-\kappa_\text{in}^2}, \\
        \tau_1 =& \frac{2\kappa_\text{ex}(g^2-\gamma^2)}{g^4 + 2g^2\gamma \kappa_\text{in} - \gamma^2(\kappa_\text{ex}^2-\kappa_\text{in}^2)},
    \end{aligned}
\end{equation}
and the difference is given by
\begin{equation} \label{eq:tau_1-tau_0}
    \tau_1-\tau_0 = \frac{2g^2\kappa_\text{ex}(\kappa_\text{ex}^2 -\kappa_\text{in}^2 -2\gamma\kappa_\text{in}-g^2)}{[g^4+2g^2\gamma \kappa_\text{in} -\gamma^2(\kappa_\text{ex}^2-\kappa_\text{in}^2)](\kappa_\text{ex}^2-\kappa_\text{in}^2)}.
\end{equation}
In the case of optimal external coupling rate $\kappa_\text{ex} = \kappa_\text{ex}^\text{opt}$ in Eq.~\eqref{eq:kappa_ex_opt}, we obtain
\begin{equation} \label{eq:tau_1-tau_0_with_Cin}
    \begin{gathered}
        \tau_0 = \frac{1}{\kappa_\text{in}}\frac{\sqrt{1+2C_\text{in}}}{C_\text{in}}, \quad \tau_1 = \frac{2C_\text{in}\kappa_\text{in}-\gamma}{\gamma \kappa_\text{in}}\frac{1}{C_\text{in}\sqrt{1+2C_\text{in}}}, \\
        \tau_1-\tau_0 = \frac{2[C_\text{in}\kappa_\text{in}-(1+C_\text{in})\gamma]}{\gamma\kappa_\text{in}C_\text{in}\sqrt{1+2C_\text{in}}}.
    \end{gathered}
\end{equation}

\subsubsection{Infidelity evaluation with Gaussian pulses}\label{app:gaussian-model}
\begin{figure}
    \centering
    \includegraphics[width=\linewidth]{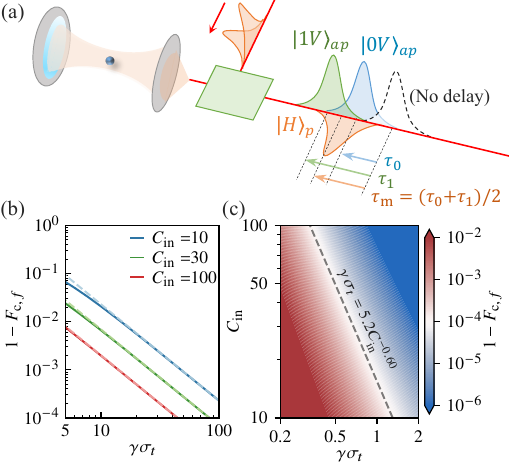}
    \caption{Mitigation of the pulse-delay effect.
    (a) Schematic of the pulse delays that depend on the qubit states of the atom and the photon.
    For the photonic qubit $\ket{0}_p = \ket{H}_p$, we introduce the calibrated delay $\tau_\mathrm{m} = (\tau_0+\tau_1)/2$ (see also Fig.~\ref{fig1}) to mitigate the qubit-dependent pulse delay effect.
    (b) CAPS gate infidelity $1-F_{c,f}$ as a function of the pulse width $\gamma \sigma_t$ for various $C_\text{in}$, before optimizing the cavity length.
    Here, we set $\kappa_\text{in}/\gamma = 0.2/3$~\cite{Hartung2024}.
    Dashed lines represent approximate results from Eq.~\eqref{app_eq:approximate_F_c_f}, agreeing well with the full calculations (solid lines) from Eq.~\eqref{S_eq:F_pro-L_f}.
    (c) CAPS gate infidelity as a function of pulse width $\gamma \sigma_t$ and internal cooperativity $C_{\text{in}}$, where the cavity length is tuned at the respective optimum, to ensure the condition given in Eq.~\eqref{eq:condition_for_kappa_in_gamma}.
    The dashed line represents the empirical criterion, $\gamma\sigma_t > 5.2 C_\text{in}^{-0.60}$, required to maintain infidelity of the CAPS gate below $10^{-4}$.
    }
    \label{fig7}
\end{figure}
To evaluate the tradeoff between speed and fidelity in the CAPS gate, we consider a canonical example in which the input mode function $f(\Delta)$ is Gaussian:
\begin{equation}
    {f}(\Delta) = \frac{1}{(\pi \sigma_{\omega}^2)^{1/4}}\exp\ab(-\frac{\Delta^2}{2\sigma_\omega^2}),
\end{equation}
Then, the mode function in time domain is written by
\begin{equation} \label{S_sec:temporal_gaussian_function}
    \begin{aligned}
        f(t) =& \frac{1}{\sqrt{2\pi}}\int \dd{\Delta} {f}(\Delta) e^{-i\Delta t} \\
        =&\frac{1}{(\pi \sigma_{t}^2)^{1/4}}\exp\ab(-\frac{t^2}{2\sigma_t^2}),
    \end{aligned}
\end{equation}
with $\sigma_t = 1/\sigma_\omega$.
For $r_j(\Delta) \simeq r_je^{i\tau_j \Delta}$, we find $\aab*{{f}_j, {f}_j} \simeq |r_j|^2$ and $\aab*{{f}, {f}_j} \simeq r_j  e^{-(\tau_j-\tau_{\text{m}})^2 \sigma_\omega^2/4}$, leading to
\begin{equation}
    \begin{aligned}
        F_{\text{pro},f} \simeq & \frac{(r^\text{opt})^2 [2+ e^{-(\tau_0-\tau_\text{m})^2\sigma_\omega^2/4} + e^{-(\tau_1-\tau_\text{m})^2\sigma_\omega^2/4}]^2}{16}, \\
        1-L_f \simeq& (r^\text{opt})^2,
    \end{aligned}
\end{equation}
where we have used $-r_0 = r_1 = r_\text{m} = r^\text{opt}$.

To mitigate the pulse-delay effect, we set $\tau_\text{m} = (\tau_0+\tau_1)/2$~\cite{Utsugi2025}, resulting in the conditional infidelity as
\begin{equation}
    \begin{aligned}
        1-{F}_{c,f} =& \frac{4}{5}\ab(1- \frac{F_{\text{pro},f}}{1-L_f})  \\
        \simeq& \frac{4}{5} \ab\{1-\ab[\frac{1+e^{-(\tau_1-\tau_0)^2 \sigma_\omega^2/16}}{2}]^2\}.
    \end{aligned}
\end{equation}
When the pulse width $\sigma_t=1/\sigma_\omega$ is sufficiently longer than the differential time delay $\tau_1-\tau_0$, corresponding to $(\tau_1-\tau_0)\sigma_\omega \ll 1$, we find
\begin{equation} \label{app_eq:approximate_F_c_f}
    1-{F}_{c,f} \simeq \frac{(\tau_1-\tau_0)^2}{20}\sigma_\omega^2 = \frac{1}{20}\ab(\frac{\tau_1-\tau_0}{\sigma_t})^2,
\end{equation}
which is shown in Fig.~\ref{fig7}(b).

\subsubsection{Optimal cavity length for pulse-delay compensation} \label{app:cavity-length-optimization}

To eliminate the atomic-state-dependent delay in Eq.~(\ref{eq:tau_1-tau_0}) by enforcing \(\tau_{0}=\tau_{1}\), the optimal external coupling rate for pulse-delay compensation is given by
\begin{equation} \label{S_eq:kappa_ex_delay}
    \kappa_{\mathrm{ex}}^{\mathrm{delay}}
    =\sqrt{\kappa_{\mathrm{in}}^{2}+2\gamma\kappa_{\mathrm{in}}+g^{2}} .
\end{equation}
To meet this condition and the reflectivity-matching requirement simultaneously, we set $\kappa_{\mathrm{ex}}^{\mathrm{delay}}=\kappa_{\mathrm{ex}}^{\mathrm{opt}}$, which yields
\begin{equation}\label{eq:length-opt-kex}
    \frac{\kappa_{\mathrm{in}}}{\gamma} = \frac{1+C_{\mathrm{in}}}{C_{\mathrm{in}}}.
\end{equation}
A practical way to satisfy Eq.~(\ref{eq:length-opt-kex}) is to adjust the cavity length, because $\kappa_{\mathrm{in}}$ scales inversely with $L_{\mathrm{cav}}$, whereas $\gamma$ and $C_\mathrm{in}$ are independent of $L_\mathrm{cav}$.
To make this dependence explicit, we express the key cavity-QED parameters in terms of $L_{\mathrm{cav}}$~\cite{Nemet2020}:
\begin{equation} \label{S_eq:parameters_depending_on_L_cav}
    g = \sqrt{\frac{v_{g}\Gamma_{\mathrm{1D}}}{L_{\mathrm{cav}}}}, ~~
    \kappa_{\mathrm{ex}} = \frac{v_{g}T_{\mathrm{ex}}}{4L_{\mathrm{cav}}},~~
    \kappa_{\mathrm{in}} = \frac{v_{g}\alpha_{\mathrm{loss}}}{4L_{\mathrm{cav}}},
\end{equation}
where $v_{g}$ is the group velocity of light, $T_{\mathrm{ex}}$ is the coupling-mirror transmittance, and  $\alpha_{\mathrm{loss}}$ is the round-trip intrinsic loss.
The emission rate into the guided mode is
\begin{equation}
    \Gamma_{\mathrm{1D}} = \frac{c}{v_g}\frac{\sigma_0}{A_\text{eff}}\gamma.
\end{equation}
As a result, the internal cooperativity is rewritten by
\begin{equation}\label{eq:cin-length}
    C_{\mathrm{in}}=\frac{g^2}{2\kappa_\mathrm{in}\gamma}
    = \frac{c}{v_{g}}\frac{\sigma_{0}}{A_{\mathrm{eff}}}\frac{2}{\alpha_{\mathrm{loss}}}
\end{equation}
which is independent of $L_{\mathrm{cav}}$.

Substituting Eqs.~(\ref{S_eq:parameters_depending_on_L_cav}) and (\ref{eq:cin-length}) into Eq.~(\ref{eq:length-opt-kex}), we find that the condition $\kappa_{\mathrm{ex}}^{\mathrm{delay}}=\kappa_{\mathrm{ex}}^{\mathrm{opt}}$ is met when the cavity length is tuned to~\cite{Utsugi2025}
\begin{equation} \label{S_eq:opt_L_cav}
    L_{\mathrm{cav}}^{\mathrm{opt}}
    = \frac{1}{1+C_{\mathrm{in}}}
      \frac{\sigma_{0}}{A_{\mathrm{eff}}}
      \frac{c}{2\gamma}.
\end{equation}
Thus, fine-tuning the cavity length offers a straightforward experimental knob for simultaneously canceling the atomic-state-dependent delay and achieving both temporal-mode and reflectivity matching.

Assuming the cavity-length optimization, we numerically evaluate the gate infidelity induced by the higher-order effects, as shown in Fig.~\ref{fig7}(c); this gives the empirical condition to realize infidelity below $10^{-4}$,
\begin{equation} \label{ap_eq:sigma_omega_condition}
    \sigma_t > 5.2 C_\mathrm{in}^{-0.60}/\gamma.
\end{equation}

\subsection{Robustness of CAPS gates}
Here, we model and quantify the response of the CAPS-gate fidelity to major imperfections expected in realistic implementations.
We consider both static deviations of the cavity parameters from the desired value due to fabrication errors, as well as random changes in the parameters arising from experimental drifts and fluctuations.
The CAPS protocol allows up to tens of percent in random, real-time fluctuations of key parameters while maintaining high-fidelity operation.
Strikingly, even greater static parameter differences between multiple atom-cavity systems are tolerated with no effect on the fidelity, thanks to the independent calibrations of atom-cavity parameters possible for passive interconnects, as we have identified in the main text.

\begin{figure}
    \centering
    \includegraphics[width=0.9\linewidth]{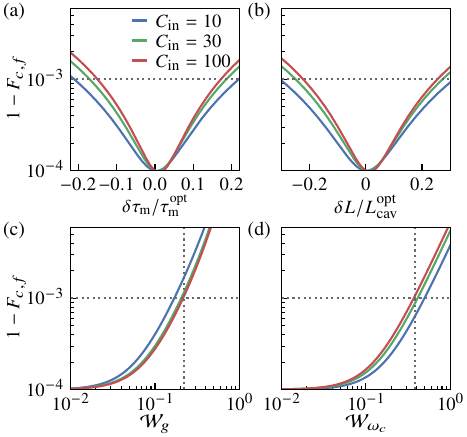}
    \caption{CAPS gate in the presence of imperfections and fluctuations.
    (a) Infidelity $1-F_{c,f}$ calculated from Eqs.~\eqref{app_eq:conditional_infidelity_with_F_pro_L} and \eqref{S_eq:F_pro-L_f} as a function of the relative deviation $\delta \tau_{\rm m}/\tau_{\rm m}^\text{opt}$ for different $C_{\rm in}$.
    (b) Effect of the static cavity-length deviation $\delta L$ from the desired value $L_\text{cav}^\text{opt}$, for example, from the fabrication error, where the cavity parameters change to $g\to g/\sqrt{1+\delta L/L_\text{cav}^\text{opt}}$ and $ \kappa_\text{ex(in)} \to \kappa_\text{ex(in)}/(1+\delta L/L_\text{cav}^\text{opt})$.
    (c) Effect due to the fluctuation of the atom-photon coupling strength $g$, where $g$ fluctuates following a Gaussian distribution around the original value $g_\text{o}$ with FWHM $g_\text{o}\mathcal{W}_g$.
    (d) Cavity-frequency jitter with FWHM $\sigma_\omega\mathcal{W}_{\omega_c}$ where $\sigma_\omega$ is the photon bandwidth which is chosen to achieve the CAPS-gate infidelity of $10^{-4}$ in the absence of fluctuation, according to Eq.~\eqref{ap_eq:sigma_omega_condition}.
    }
    \label{fig8}
\end{figure}

First, we discuss the effect of deviations in atom-photon coupling $g$ among the cavities used for remote entanglement generation.
Such an effect is detrimental in the remote entanglement generation protocol using photon emission and two-photon interference, since the difference in cavity parameters degrades the indistinguishability of the emitted photons from different setups.
Consider two atom-cavity systems operating sequential CAPS networking, where the first cavity has the atom-photon coupling $g$ with internal cooperativity $C_\mathrm{in}$, and the second cavity has $g^\prime$ and $C_\mathrm{in}^\prime$.
For each cavity, we independently set the outcoupling rate $\kappa_\mathrm{ex}$ to satisfy Eq.~\eqref{eq:kappa_ex_opt}: this is possible in situ for various cavity implementations, such as the nanofiber cavity with precise thermal tuning capability of mirror reflectivity~\cite{Kato2019}, the fiber-taper-coupled microresonator with finely tuned taper-resonator distance~\cite{Spillane2003, Volz2014, Bechler2018} or the free-space cavity with an output coupler placed outside the vacuum chamber~\cite{Shadmany2025}.
With appropriate tuning of the HWP angle $\theta_r$ and delay line $\tau_\text{m}$ for each device, reflectivity mismatch and pulse delay errors are eliminated independently.
To investigate the requirement for the calibration of the delay line, we evaluate the increase of infidelity arising from the imperfect delay line $\tau_\text{m}$, as shown in Fig.~\ref{fig8}(a); with the pulse length $\sigma_t$ to be at the right-hand side of Eq.~\eqref{ap_eq:sigma_omega_condition}, we find that the infidelity is suppressed below $10^{-3}$ up to more than $10\%$ deviation of the delay line.

Further controlling the cavity length $L_\mathrm{cav}$ independently, e.g., by fiber-Bragg-grating placement for the nanofiber cavity~\cite{Sunami2025,Horikawa2025} or setting the voltages for the piezoelectric adjuster for free-space cavities~\cite{Shadmany2025}, and setting the single-photon pulse width $\sigma_t$ to satisfy Eq.~\eqref{ap_eq:sigma_omega_condition} for both cavities, then the overall infidelity is suppressed to $10^{-4}$ per CAPS gate, independent of the fractional differences of $g$ and $g^\prime$.
Figure~\ref{fig8}(b) shows the CAPS-gate infidelity as a function of fractional deviation $\delta L$ from the optimal cavity length $L_\mathrm{cav}^\mathrm{opt}$.
The results indicate that maintaining the infidelity below $10^{-3}$ requires fractional length precision of $\lesssim 0.2$, i.e., 20\% deviation of the cavity length is permitted for high-fidelity operation.
This demonstrates notable tolerance of the CAPS gate to the fabrication errors.

In Fig.~\ref{fig8}(c), we plot the effect of random fluctuations in the atom-photon coupling $g$ on the conditional infidelity of the CAPS gate, with other parameters fixed.
This quantifies the robustness of the CAPS gate to real-time and post-installation fluctuations arising, for example, from the finite temperature of the trapped atoms and fluctuations of the spatial cavity mode.
In our simulation, the coupling $g$ follows a Gaussian distribution with a full width at half maximum (FWHM) of $\mathcal{W}_g$ in units of $g$, i.e., fractional fluctuation with FWHM $\mathcal{W}_g$.
According to Fig.~\ref{fig8}(c), nearly 20\% fractional fluctuation of $g$ is allowed while maintaining the CAPS-gate infidelity below $10^{-3}$.

Finally, we evaluate the performance of the CAPS gate under fluctuations in the cavity resonance frequency $\omega_c$, which we denote as $\delta \omega_c$ arising, for example, from cavity lock jitter.
Here, $\omega_c$ fluctuates around its desired frequency following a Gaussian distribution with FWHM $\mathcal{W}_{\omega_c}$ in units of the photon bandwidth $\sigma_\omega~(=1/\sigma_t)$, which is set according to Eq.~\eqref{ap_eq:sigma_omega_condition}.
This fluctuation not only shifts the cavity response in Eq.~\eqref{eq:freq-dependent-reflectance} as $\Delta \to \Delta - \delta \omega_c$ but also detunes the resonance between the cavity and the atom [see Eq.~\eqref{app_eq:freq-dep-reflection} for the response function including the shift of the cavity resonance].
Figure~\ref{fig8}(d) shows that the CAPS gate is highly robust against this error, with up to $\approx 10\%$ jitter resulting in a negligible increase of infidelity, while nearly $40\%$ fluctuation is allowed for the total infidelity of $10^{-3}$.

\section{CAPS-based memory loading} \label{s_sec:memory_loading}
Here, we analyze CAPS-based memory loading by following the discussion in Ref.~\cite{Raymer2024} with revisions made primarily to simplify the notations, and derive an operator that characterizes the error arising from the frequency dependence of the reflection amplitudes.
The atom is initially prepared in $\ket{+}_a$, and the photonic qubit is $\ket{\psi}_p = \alpha\ket{0}_p + \beta\ket{1}_p~(|\alpha|^2+|\beta|^2 = 1)$, without considering the photonic frequency spectrum.
In the memory loading scheme, we finally measure the photonic qubit state, which allows us to postselect the trajectory without photon loss.
Thus, in what follows, we only track it, where $\hat{G}_0$ represents the action of the CAPS gate.
Applying the CAPS gate to the initial state $\ket{+}_a(\alpha\ket*{0;{f}}_p + \beta\ket*{1;{f}}_p)$ yields
\begin{equation}
    \begin{aligned}
        &\alpha\ket{+}_a r_\text{m}\ket*{0;{f}}_p + \frac{\beta}{\sqrt{2}} \ab(\ket{0}_a \ket*{1;{f}_0}_p + \ket*{1}_a \ket*{1; {f}_1}_p), \\
        &= \ket{+}_a \ab(\alpha r_\text{m} \ket*{0;{f}}_p + \beta \ket*{1; {f}_+}_p) - \beta\ket{-}_a \ket*{1;{f}_-}_p),
    \end{aligned}
\end{equation}
where ${f}_{\pm}(\Delta) = [f_1(\Delta) \pm f_0(\Delta)]/2$.
Applying the Hadamard gates $\hat{H}_a\hat{H}_p$ results in
\begin{equation}
    \ket{\phi}_{ap} = \ket{0}_a \ab(\alpha r_\text{m}\ket*{+;{f}}_p + \beta \ket*{-; {f}_+}_p) - \beta\ket{1}_a \ket*{-; {f}_-}_p),
\end{equation}
which reduces to $\hat{Z}_a\ket{\psi}_a \ket*{0;{f}}_p + \ket{\psi}_a \ket*{1;{f}}_p$ in the ideal case, $-{r}_0(\Delta)={r}_1(\Delta) = r_\text{m} = 1$ and $\tau_\text{m}=0$.
For a detector having a flat frequency response, the measurement operator of detecting the photonic qubit $j\in\{0,1\}$ is given by
\begin{equation}
    \hat{\Pi}_j = \int\dd{\Delta}\hat{a}_j^\dagger(\Delta) \ketbra{\varnothing}[_p]{\varnothing}\hat{a}_j(\Delta).
\end{equation}
From the relation
\begin{equation}
    _p\bra{\varnothing}\hat{a}_j(\Delta)\ket{\phi}_{ap} = \frac{f(\Delta)}{\sqrt{2}} \hat{E}_a(\Delta) \hat{Z}_a^{1+j}\ket{\psi}_a,
\end{equation}
where
\begin{equation}\label{S_eq:memoryloading}
    \hat{E}(\Delta) = r_\text{m}\ketbra{0}{0} + e^{-i\tau_\text{m}\Delta}[{r}_-(\Delta)\ket{1} - {r}_+(\Delta)\ket{0}]\bra{1},
\end{equation}
and ${r}_\pm(\Delta) = [{r}_1(\Delta) \pm {r}_0(\Delta)]/2$, we obtain the density operator of the atom after measurement $j\in\{0,1\}$ as
\begin{equation}
    \begin{aligned}
        \hat{\rho}^{(j)}_\text{load} =& \frac{\Tr_p[\hat{\Pi}_j \ketbra{\phi}[_{ap}]{\phi}]}{\Tr[\hat{\Pi}_j \ketbra{\phi}[_{ap}]{\phi}]} \\
        =& \frac{1}{P_\text{load}^{(j)}} \int \dd{\Delta} \frac{\vab{f(\Delta)}^2}{2} \hat{E}_a(\Delta)\hat{Z}_a^{1+j} \ketbra{\psi}[_a]{\psi} \hat{Z}_a^{1+j}\hat{E}_a^\dagger(\Delta),
    \end{aligned}
\end{equation}
where
\begin{equation}
    P_\text{load}^{(j)} = \int \dd{\Delta} \frac{\vab{{f}(\Delta)}^2}{2}\,_a\braket*[3]{\psi}{\hat{Z}_a^{1+j}\hat{E}_a^\dagger(\Delta)\hat{E}_a(\Delta)\hat{Z}_a^{1+j}}{\psi}_a,
\end{equation}
represents the detection probability.
Here, $\hat{Z}^{1+j} \ket{\psi}$ is the ideal final state, and the operator $\hat{E}(\Delta)$ represents the error induced by the frequency dependence of the reflection amplitudes.

\section{Cavity-assisted single-photon and atom-photon entanglement generation} \label{s_sec:photon_generation}
Here, we develop a theoretical framework for the generation of a single photon and atom-photon entanglement using cavity-QED systems, which serve as core functionalities of the sequential CAPS and emission-CAPS networking.
We begin by analyzing the emission of single photons from a $\Lambda$-type atomic system and characterizing their temporal properties in Appendix~\ref{app:photon_source_cavity_QED}.
Building on this foundation, we then consider the generation of atom-photon entangled states through polarization-selective cavity coupling in Appendix~\ref{S_seq:atom_photon_entanglement}.

\subsection{Cavity-assisted single-photon generation} \label{app:photon_source_cavity_QED}
We numerically evaluate the single photon generation with a $\Lambda$-type three-level system coupled to a cavity, as shown in Fig.~\ref{fig2}(a).
The atom is initially prepared in $\ket{u}_a$ at time $t=t_\text{i}$.
The Hamiltonian of the system is given by
\begin{equation}
    \hat{H}_s(t) = \Omega(t) (\ketbra{e}[_a]{u} + \ketbra{u}[_a]{e}) + g(\ketbra{e}[_a]{g}\hat{c} + \ketbra{g}[_a]{e}\hat{c}^\dagger),
\end{equation}
and the atomic decay and the internal cavity loss are denoted by the following Lindblad operators:
\begin{equation} \label{S_eq:Lindblad_ops_photon_gen}
    \begin{aligned}
        \hat{L}_1 =& \sqrt{2\kappa_\text{in}}\hat{c}, \\
        \hat{L}_2 =& \sqrt{2p_\text{br}\gamma}\ketbra{u}[_a]{e}, \\
        \hat{L}_3 =& \sqrt{2(1-p_\text{br})\gamma}\ketbra{g}[_a]{e},
    \end{aligned}
\end{equation}
where $p_\text{br}$ denotes the branching ratio of the atomic decay to the initial state $\ket{u}_a$.
The cavity couples to the output mode at rate $\kappa_\text{ex}$.

For $p_\text{br} = 0$, where the spontaneous emission at rate $\gamma$ always leads to failure of the photon generation, the atom-cavity system probabilistically emits a pure photon.
In contrast, for $p_\text{br} > 0$, the atomic decay $\hat{L}_2$ resets the atom in the initial state $\ket{u}_a$, thereby restarting the photon generation process.
This reexcitation process results in the photon emission with a distorted wave packet~\cite{Maraner2020, Tanji2024, Kikura2025_high_purity}.
The generated photonic state is given by~\cite{Maraner2020,Kikura2025}
\begin{equation} \label{S_eq:photonic_state_with_reexcitation}
    \begin{aligned}
        \hat{\varrho} =& \ketbra*{n=1;\psi_{t_\text{i}}}[_p]{n=1;\psi_{t_\text{i}}}  \\
        &+ \int_{t_\text{i}}^\infty \dd{s} r(s) \ketbra{n=1;\psi_s}[_p]{n=1;\psi_s} \\
        &+ (1-P_\text{gen})\ketbra{\varnothing}[_p]{\varnothing},
    \end{aligned}
\end{equation}
where $\ket{n=1;\psi_s}_p~(s\geq t_\text{i})$ is the unnormalized single-photon state corresponding to a trajectory in which the atomic decay $\hat{L}_2$ occurs at $t=s$ and does not occur for $t>s$.
The state $\ket*{n=1;\psi_{t_\text{i}}}$ represents the trajectory without the decay $\hat{L}_2$, and the function  $r(s)$ denotes the decay rate associated with  $\hat{L}_2$ at $t=s$.
Then, the photon generation probability is given by
\begin{equation}
    \begin{aligned}
        P_\text{gen} =&\ _p\braket*{n=1;\psi_{t_\text{i}}}{n=1;\psi_{t_\text{i}}}_p \\
        &+ \int_{t_\text{i}}^\infty \dd{s} r(s)\ _p\braket{n=1;\psi_s}{n=1;\psi_s}_p.
    \end{aligned}
\end{equation}

To characterize the photonic state in Eq.~\eqref{S_eq:photonic_state_with_reexcitation}, we use the temporal autocorrelation function~\cite{Fabre2020},
\begin{equation} \label{app_eq:def_g^1}
    g^{(1)}(t,t^\prime) \coloneqq \Tr[\hat{a}^\dagger(t) \hat{a}(t^\prime)\hat{\varrho}],
\end{equation}
where
\begin{equation}
    \hat{a}(t) = \frac{1}{\sqrt{2\pi}}\int \dd{\Delta} \hat{a}(\Delta) e^{-i\Delta t}
\end{equation}
is the instantaneous annihilation operator, which satisfies $[\hat{a}(t), \hat{a}^\dagger(t^\prime)] = \delta(t-t^\prime)$.
We rewrite the photonic state with the autocorrelation function in Eq.~\eqref{app_eq:def_g^1} as follows:
\begin{equation} \label{S_eq:reexcitation_photon_with_g^1}
    \begin{aligned}
        \hat{\varrho} =& \iint \mathrm{d}t \dd{t^\prime} g^{(1)}(t,t^\prime) \hat{a}^\dagger(t^\prime) \ketbra{\varnothing}[_p]{\varnothing}\hat{a}(t) \\
        & + (1-P_\text{gen}) \ketbra{\varnothing}[_p]{\varnothing},
    \end{aligned}
\end{equation}
where
\begin{equation} \label{S_eq:g^1_for_reexcitation}
    g^{(1)}(t,t^\prime) =  \psi_{t_\text{i}}^\ast(t) \psi_{t_\text{i}}(t^\prime) + \int_{t_\text{i}}^\infty \dd{s}r(s) \psi_{s}^\ast(t) \psi_{s}(t).
\end{equation}

To quantitatively evaluate the photonic state, we simulate the dynamics of the local atom-cavity system, treating the desired mode as part of the environment.
In this case, the external coupling is also expressed by the Lindblad operator, $\hat{L}_0 = \sqrt{2\kappa_\text{ex}}\hat{c}$, and the system evolves according to the master equation as follows:
\begin{equation} \label{S_eq:master_eq_for_photon_source}
    \odv{\hat{\rho}}{t} = -i[\hat{H}_s(t), \hat{\rho}] + \sum_{j=0}^3 \ab(\hat{L}_j \hat{\rho}\hat{L}_j^\dagger -\frac{1}{2}\{\hat{L}_j^\dagger\hat{L}_j, \hat{\rho}\}).
\end{equation}
We denote the solution with the dynamical map, $\hat{\rho}(t) = \Lambda(t;t_0)[\hat{\rho}(t_0)]$~\cite{Campaioli2024}.
This map gives the autocorrelation function of the emitted photon as follows~\cite{Kiilerich2019, Kiilerich2020}:
\begin{equation}
    g^{(1)}(t,t^\prime) = \Tr\ab[ \hat{L}_0^\dagger \Lambda(t; t^\prime)[\hat{L}_0 \hat{\rho} (t^\prime)]] \quad (t\geq t^\prime),
\end{equation}
providing the full information of $g^{(1)}(t,t^\prime)$, since $g^{(1)}(t^\prime, t) = [g^{(1)}(t,t^\prime)]^\ast$ by definition of Eq.~\eqref{app_eq:def_g^1}.
We numerically calculate this and obtain the temporal autocorrelation function with QuTiP~\cite{Lambert2026}.
Note that the autocorrelation function can be experimentally accessed via homodyne measurement~\cite{Morin2013}.

\subsection{Atom-photon entanglement generation} \label{S_seq:atom_photon_entanglement}
As an extension of the single-photon generation discussed in Appendix~\ref{app:photon_source_cavity_QED}, we further evaluate the atom-photon entanglement generation.
We consider the typical level structure of the entanglement generation~\cite{Reiserer2015, Hartung2024} [Fig.~\ref{fig3}(a)], where the transition $\ket{0}_a \leftrightarrow \ket{e}_a$ ($\ket{1}_a \leftrightarrow \ket{e}_a$) is coupled to the left (right) circularly polarized cavity mode.
For simplicity, we consider that the two cavity modes couple to the atom at the same coupling strength $g$.
The Hamiltonian is given by
\begin{equation}
    \begin{aligned}
        \hat{H}_s(t) =& \Omega(t) (\ketbra{e}[_a]{u} + \ketbra{u}[_a]{e}) \\
        & + g \sum_{j=0,1}(\ketbra{e}[_a]{j}\hat{c}_j + \ketbra{j}[_a]{e}\hat{c}_j^\dagger),
    \end{aligned}
\end{equation}
where $\hat{c}_{0(1)}$ is the annihilation operator of the left (right) circularly polarized mode.
The Lindblad operators are given by $(j \in \{0,1\})$:
\begin{equation}
    \begin{alignedat}{2}
        \hat{L}_{0j} =& \sqrt{2\kappa_\text{ex}}\hat{c}_j \\
        \hat{L}_{1j} =& \sqrt{2\kappa_\text{in}}\hat{c}_j \\
        \hat{L}_2 =& \sqrt{2p_\text{br}\gamma} \ketbra{u}[_a]{e}, \\
        \hat{L}_{3j} =& \sqrt{(1-p_\text{br})\gamma} \ketbra{j}[_a]{e} \\
    \end{alignedat}
\end{equation}
The desired atom-photon entangled state is
\begin{equation}
    \ket*{\Phi^+; f}_{ap} = \frac{\ket{0}_a\ket*{0;f}_p + \ket{1}_a\ket*{1;f}_p}{\sqrt{2}},
\end{equation}
followed by the photon passing through the waveplate.
As in the case of the single-photon generation, the atomic decay to $\ket{u}_a$ causes the generation of the atom-photon entangled state in the distorted wave packet, resulting in the mixed state as follows~\cite{Krutyanskiy2023}:
\begin{equation} \label{S_eq:photonic_state_with_reexcitation_atomphoton}
    \begin{aligned}
        \hat{\rho}_{ap} =& \ketbra*{\Phi^+; \psi_{t_\text{i}}}[_{ap}]{\Phi^+; \psi_{t_\text{i}}}  \\
        &+ \int_{t_\text{i}}^\infty \dd{s} r(s) \ketbra*{\Phi^+; \psi_{s}}[_{ap}]{\Phi^+; \psi_{s}} \\
        &+ (1-P_\text{gen})\hat{\rho}_{a\varnothing},
    \end{aligned}
\end{equation}
where $\hat{\rho}_{a\varnothing}$ represents the failure of the photon generation.
In this case, the autocorrelation function in Eq.~\eqref{S_eq:g^1_for_reexcitation} is given by
\begin{equation}
    g^{(1)}(t,t^\prime) = \sum_{j=0,1} \Tr[\hat{a}_j^\dagger(t) \hat{a}_j(t^\prime) \hat{\rho}_{ap}],
\end{equation}
which can be calculated from the dynamics of the atom-cavity system as follows:
\begin{equation}
    g^{(1)}(t,t^\prime) = \sum_{j=0,1}\Tr\ab[ \hat{L}_{0j}^\dagger \Lambda(t; t^\prime)[\hat{L}_{0j} \hat{\rho} (t^\prime)]] \quad (t\geq t^\prime).
\end{equation}
Note that we set $\Omega(t)$ by replacing $(g, \kappa_\text{ex}, \kappa_\text{in})$ with $(2g, 2\kappa_\text{ex}, 2\kappa_\text{in})$ in the analytical expression of $\Omega(t)$ for the single-photon generation~\cite{Utsugi2022}, so that the generated wave packet is close to the desired Gaussian function.
This adjustment accounts for the two cavity-coupling pathways involved in the entanglement generation protocol.

The theoretical framework developed here is subsequently utilized to evaluate heralded entanglement generation in Appendix~\ref{ap_sec:HEG}.

\section{Heralded remote entanglement generation} \label{ap_sec:HEG}
Here, we analyze remote entanglement generation protocols based on CAPS gates, focusing on two representative network configurations: sequential CAPS and emission-CAPS networking.
For later convenience, we refer to the two atom-cavity systems as Alice (A) and Bob (B), between which entanglement is established.
In Appendix~\ref{s_sec:type-II-networking_with_photon}, we consider the sequential CAPS networking where single photons are supplied by an external source and sequentially interact with two atom-cavity systems to generate heralded entanglement.
In Appendix~\ref{app_seq:type-II-bellpair}, we propose a heralded entanglement generation (HEG) protocol that uses an external entangled photon-pair source and CAPS-based memory loading, enabling improved performance in high-loss regimes such as satellite-based links.
In Appendix~\ref{s_sec:type-III}, we analyze the emission-CAPS networking, which combines atom-photon entanglement generation at one node with CAPS-based memory loading at the other, eliminating the need for external photon sources while maintaining high fidelity and success probability.
In Appendix~\ref{s_sec:type-I}, we further consider the photon-interference-based networking with imperfect atom-photon entanglement as a reference, and show the infidelity arising from photon impurity.

\subsection{Sequential CAPS networking with single-photon sources} \label{s_sec:type-II-networking_with_photon}

To evaluate the performance of sequential CAPS networking, we derive two key metrics: conditional fidelity and success probability for the protocol in which sequential CAPS gates and a final photonic measurement are used to generate entanglement between Alice (A) and Bob (B) assisted by an ancilla photon.
Specifically, for the atomic-qubit input state $\ket{+}^\text{A}\ket{+}^\text{B}$ and a photon initially in the state $\ket*{+;{f}}_p$, the (unnormalized) premeasurement state is obtained using the CAPS gate operator $\hat{G}_{0,f}$ in Eq.~(\ref{S_eq:mode_incorporated_G_0}):
\begin{equation}
    \begin{aligned}
        &\hat{G}_{0,f}^\text{B} \hat{X}_p \hat{G}_{0,f}^\text{A} \ket{+}^\text{A}\ket{+}^\text{B}\ket*{+;{f}}_p \\
        &= \frac{1}{2\sqrt{2}} \Big[ \ket{00}(r_\text{m}^\text{A} \ket*{1;{f}_0^\text{B}}_p +r_\text{m}^\text{B}\ket*{0;{f}_0^\text{A}}_p) \\
        & \hspace{1.5cm} + \ket{11} (r_\text{m}^\text{A} \ket*{1;{f}_1^\text{B}}_p +r_\text{m}^\text{B}\ket*{0;{f}_1^\text{A}}_p) \\
        & \hspace{1.5cm} + \ket{01} (r_\text{m}^\text{A} \ket*{1;{f}_1^\text{B}}_p +r_\text{m}^\text{B}\ket*{0;{f}_0^\text{A}}_p) \\
        & \hspace{1.5cm} + \ket{10} (r_\text{m}^\text{A} \ket*{1;{f}_0^\text{B}}_p +r_\text{m}^\text{B}\ket*{0;{f}_1^\text{A}}_p) \Big] \\
        &\eqqcolon \ket{\psi}_\text{pre}
    \end{aligned}
\end{equation}
Then, we measure the photonic qubit in $X$ basis.
The final two-atom state, conditioned on the measurement outcome $j\in\{0,1\}$ can be derived using the following relation:
\begin{equation}
    \begin{aligned}
        &_p\bra{\varnothing}\hat{a}_j(\Delta)\hat{H}_p \ket{\psi}_\text{pre}  \\
        &= \frac{{f}(\Delta)}{4} \Big\{[\mathrm{r}^\text{A}_0(\Delta)r_\text{m}^\text{B} +(-1)^j r_\text{m}^\text{A} \mathrm{r}^\text{B}_0(\Delta) ] \ket{00} \\
        &\hspace{1.5cm} + [\mathrm{r}^\text{A}_1(\Delta)r_\text{m}^\text{B} +(-1)^j r_\text{m}^\text{A} \mathrm{r}^\text{B}_1(\Delta)]\ket{11} \\
        & \hspace{1.5cm} + [\mathrm{r}^\text{A}_0(\Delta)r_\text{m}^\text{B} +(-1)^j r_\text{m}^\text{A} \mathrm{r}^\text{B}_1(\Delta)]\ket{01} \\
        & \hspace{1.5cm} + [\mathrm{r}^\text{A}_1(\Delta)r_\text{m}^\text{B} +(-1)^j r_\text{m}^\text{A} \mathrm{r}^\text{B}_0(\Delta)]\ket{10} \Big\} \\
        &\eqqcolon \frac{-f(\Delta)}{\sqrt{2}} \ket*{\Upsilon^{(j)}(\Delta)},
    \end{aligned}
\end{equation}
where $\mathrm{r}_j^\text{q}(\Delta) = e^{-i\tau_\text{m}^\text{q}\Delta} {r}_j^\text{q}(\Delta)$ and $\ket{ij} = \ket{i}^\text{A}\ket{j}^\text{B}$.
From this, we obtain the post-measurement density operator of the two atoms as
\begin{equation}
    \begin{aligned}
        \hat{\rho}^{(j)}_\text{cc} =& \frac{\Tr_p[\hat{\Pi}_j  \hat{H}_p\ketbra{\psi}[_\text{pre}]{\psi}\hat{H}_p]}{\Tr[\hat{\Pi}_j \hat{H}_p\ketbra{\psi}[_\text{pre}]{\psi}\hat{H}_p ]} \\
        =& \frac{1}{P^{(j)}_\text{cc}} \int \dd{\Delta} \frac{\vab{f(\Delta)}^2}{2} \ketbra*{\Upsilon^{(j)}(\Delta)}{\Upsilon^{(j)}(\Delta)},
    \end{aligned}
\end{equation}
where
\begin{equation}
    P^{(j)}_\text{cc} = \int \dd{\Delta} \frac{\vab{f(\Delta)}^2}{2} \braket*{\Upsilon^{(j)}(\Delta)}{\Upsilon^{(j)}(\Delta)}
\end{equation}
is the probability of obtaining the measurement outcome $j$.

For the ideal, \textit{lossless} case where all reflection coefficients satisfy $-\mathrm{r}_0^{\text{q}}(\Delta)= \mathrm{r}_1^{\text{q}}(\Delta) = r_\text{m}^{\text{q}} = 1$ for $\text{q} \in \{\text{A}, \text{B}\}$,
we find that $\ket*{\Upsilon^{(0)}(\Delta)} = \ket*{\Phi^-}$ and $\ket*{\Upsilon^{(1)}(\Delta)} = \ket*{\Psi^-}$, where the Bell states are
\begin{equation}
    \ket*{\Phi^{\pm}} = \frac{\ket{00} \pm \ket{11}}{\sqrt{2}}, \quad \ket*{\Psi^{\pm}} = \frac{\ket{01} \pm \ket{10}}{\sqrt{2}}.
\end{equation}
In realistic scenarios, however, deviations from the ideal parameters lead to mixed output states, and the fidelity of the resulting entanglement must be evaluated accordingly.
Consequently, the conditional fidelity and the total success probability are given by:
\begin{equation} \label{app_ex:type-II-networking_measures}
    \begin{aligned}
        F_\text{cc} =& \frac{P^{(0)}_\text{cc} \braket*[3]{\Phi^{-}}{\hat{\rho}^{(0)}}{\Phi^{-}} + P^{(1)}_\text{cc} \braket*[3]{\Psi^{-}}{\hat{\rho}^{(1)}}{\Psi^{-}}}{P^{(0)}_\text{cc}+ P^{(1)}_\text{cc} }, \\
        P_\text{cc} =& P^{(0)}_\text{cc}+ P^{(1)}_\text{cc}.
    \end{aligned}
\end{equation}

\subsubsection{Robustness against the inhomogeneity of two systems}
Here, we consider the robustness of the protocol against variations between the two atom-cavity systems.
Specifically, differences in the atom-photon coupling strength
$g$ lead to distinct optimal cavity reflectivities, i.e.,  $r^\mathrm{opt, A}\neq r^\mathrm{opt, B}$.
From the relations, $r_0^\text{q}(\Delta) = -r^\text{opt,q} + \mathcal{O}(\Delta^2)$ and $r_1^\text{q}(\Delta) = r^\text{opt,q} + \mathcal{O}(\Delta^2)$, with the calibrated delay line $\tau_\text{m}$, we derive
\begin{equation}
    \begin{aligned}
         \ket*{\Upsilon^{(j)}(\Delta)} =& \frac{r^\text{opt,A}r_\text{m}^\text{B} + (-1)^jr_\text{m}^\text{A}r^\text{opt,B}}{2} \ket{\Phi^-} \\
        & +\frac{r^\text{opt,A}r_\text{m}^\text{B} - (-1)^jr_\text{m}^\text{A}r^\text{opt,B}}{2}\ket{\Psi^-} + \mathcal{O}(\Delta^2).
    \end{aligned}
\end{equation}
This means that the condition $r^\text{opt,A}r_\text{m}^\text{B} = r_\text{m}^\text{A}r^\text{opt,B}$ yields the ideal Bell states up to first order in $\Delta$.
To ensure this condition is met, we adjust the mirror reflectivities as follows:
\begin{equation} \label{app_eq:mirror_reflectivities_for_inhomogeneity}
    \begin{cases}
        r_\text{m}^\text{A} = 1, r_\text{m}^\text{B} = r^\text{opt,B}/r^\text{opt,A} \quad \mathrm{if}~~r^\text{opt,A} \geq r^\text{opt,B},\\
        r_\text{m}^\text{B} = 1, r_\text{m}^\text{A} = r^\text{opt,A}/r^\text{opt,B} \quad \mathrm{if}~~r^\text{opt,A} \leq r^\text{opt,B},
    \end{cases}
\end{equation}
which leads to a success probability of $[\min(r^\text{opt,A},r^\text{opt,B})]^2$.

\subsubsection{Photon in a mixed state}\label{app:mixed-state-photon}
So far, we have assumed that the input photon is in a pure state.
In practice, however, a realistic photon source will emit a photon in a mixed state due to, e.g., experimental imperfections or fundamental limitations of the generation scheme.
We now extend the above analysis to address this case, where the input photonic state is modeled as a statistical mixture of single-photon states~\cite{Fabre2020}.
The input photon in a mixed state is given by
\begin{equation} \label{S_eq:mixed_photon_state}
    \hat{\varrho} = \sum_l p_l \ketbra*{+;{u}_l}[_p]{+;{u}_l} + \ab(1-\sum_l p_l)\ketbra{\varnothing}[_p]{\varnothing},
\end{equation}
where $\sum_l p_l \leq 1$, and ${u}_l(\Delta)$ are the mode functions.
For the sequential CAPS networking with the incoming photon in Eq.~\eqref{S_eq:mixed_photon_state}, we straightforwardly expand the above results by replacing $|f(\Delta)|^2$ with $\sum_lp_l |u_l(\Delta)|^2$.
For this photonic state, the autocorrelation function is given by
\begin{equation}
    g^{(1)}(t, t^\prime) =\sum_l p_l u_l^\ast(t) u_l(t^\prime).
\end{equation}
Thus, given $g^{(1)}(t, t^\prime)$ as the full characterization for the mode distribution of the photon, the fidelity and success probability can be calculated using the following relation for an arbitrary function ${h}(\Delta)$:
\begin{equation}
    \begin{aligned}
        & \int d\Delta \sum_l p_l|{u}_l(\Delta)|^2 {h}(\Delta) \\
        &=  \iint \mathrm{d}t\dd{t^\prime} g^{(1)}(t, t^\prime) \frac{1}{2\pi} \int \dd{\Delta} {h}(\Delta) e^{-i \Delta(t-t^\prime)}.
    \end{aligned}
\end{equation}

\subsection{CAPS networking with photon-pair sources} \label{app_seq:type-II-bellpair}
\begin{figure}
    \centering
    \includegraphics{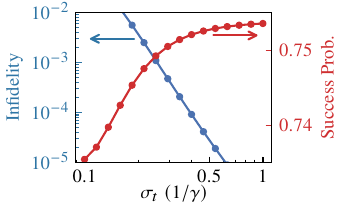}
    \caption{Infidelity and success probability for the case where the two entangled photons are in the Gaussian wave packet with $\sigma_t$ and two parties have identical systems with $C_\text{in} = 100$.}
    \label{fig9}
\end{figure}
Here, we consider the HEG protocol with entangled photon-pair sources, in which a photonic Bell state is loaded into the atomic qubits of Alice and Bob (see also Ref.~\cite{Ji2025} that proposes an efficient repeater protocol leveraging this).
First, we prepare the photonic Bell state,
\begin{equation}
    \ket{\Psi^+;f^\text{A}, f^\text{B}}_p = \frac{\ket*{0;f^\text{A}}_p\ket*{1;f^\text{B}}_p + \ket*{1;f^\text{A}}_p\ket*{0;f^\text{B}}_p}{\sqrt{2}},
\end{equation}
by, e.g., spontaneous parametric downconversion (SPDC) or quantum emitters.
Upon obtaining the measurement outcome $(j^\text{A}, j^\text{B})$ during memory loading at Alice and Bob, described by the memory-loading operator $\hat{E}(\Delta)$ in Eq.~\eqref{S_eq:memoryloading} with the ideal loaded state given by $[\ket{01} + (-1)^{j^\text{A}-j^\text{B}}\ket{10}]/\sqrt{2}$, the atomic-qubit pair is projected onto
\begin{equation}
    \begin{aligned}
        &\frac{\hat{E}^\text{A}(\Delta^\text{A})}{\sqrt{2}} \frac{\hat{E}^\text{B}(\Delta^\text{B})}{\sqrt{2}} (\hat{Z}^\text{A})^{1+j^\text{A}} (\hat{Z}^\text{B})^{1+j^\text{B}} \ket{\Psi^+} \\
        &= \frac{r_\text{m}^\text{A} \mathrm{r}_-^\text{B}(\Delta^\text{B}) \ket{01} + (-1)^{j^\text{A}-j^\text{B}} r_\text{m}^\text{B} \mathrm{r}_-^\text{A}(\Delta^\text{A}) \ket{10}}{2\sqrt{2}} \\
        &\hspace{0.5cm}- \frac{r_\text{m}^\text{A} \mathrm{r}_+^\text{B}(\Delta^\text{B}) + (-1)^{j^\text{A}-j^\text{B}} r_\text{m}^\text{B} \mathrm{r}_+^\text{A}(\Delta^\text{A}) }{2\sqrt{2}}\ket{00} \\
        &\eqqcolon \ket*{\Phi^{(j^\text{A},j^\text{B})}(\Delta^\text{A},\Delta^\text{B})},
    \end{aligned}
\end{equation}
where $\mathrm{r}_\pm(\Delta) = e^{-i\tau_\text{m}\Delta} {r}_{\pm}(\Delta)$, and we have neglected a global phase.
Thus, the loaded atomic-qubit state is given by
\begin{equation}
    \hat{\rho}_{\text{cc}^\prime}^{(j^\text{A}, j^\text{B})} = \frac{\mathbb{E}\ab[\ketbra*{\Phi^{(j^\text{A},j^\text{B})}(\Delta^\text{A},\Delta^\text{B})}{\Phi^{(j^\text{A},j^\text{B})}(\Delta^\text{A},\Delta^\text{B})}]}{P_{\text{cc}^\prime}^{(j^\text{A},j^\text{B})}},
\end{equation}
where the symbol $\mathbb{E}$ is defined for a two-variable function $h(\Delta^\text{A},\Delta^\text{B})$ as
\begin{equation}
    \begin{aligned}
        &\mathbb{E}[{h}(\Delta^\text{A},\Delta^\text{B})] \\
        &= \iint \text{d}{\Delta^\text{A}}\dd{\Delta^\text{B}} |{f}^\text{A}(\Delta^\text{A})|^2|{f}^\text{B}(\Delta^\text{B})|^2 h(\Delta^\text{A},\Delta^\text{B}),
    \end{aligned}
\end{equation}
and
\begin{equation}
    P_{\text{cc}^\prime}^{(j^\text{A},j^\text{B})} = \mathbb{E}\ab[\|\ket*{\Phi^{(j^\text{A},j^\text{B})}(\Delta^\text{A},\Delta^\text{B})}\|^2]
\end{equation}
is the success probability of the remote entanglement generation conditioned on the detection outcome $(j^\text{A}, j^\text{B})$.
From these expressions, we readily calculate the fidelity and the total success probability, demonstrating an infidelity around $10^{-3}$ for a pulse width satisfying $\gamma\sigma_t\gtrsim 0.25$ as shown in Fig.~\ref{fig9}.

\subsubsection{Robustness against the inhomogeneity of two systems}
As in Appendix~\ref{s_sec:type-II-networking_with_photon}, we analyze the protocol's robustness to system asymmetries, focusing on how variations in the atom-photon coupling $g$ between two cavities lead to differing optimal reflectivities $r^\mathrm{opt, A}\neq r^\mathrm{opt, B}$.
In the long-pulse limit, where the detuning dependence is negligible and $\ket*{\Phi^{(j^\text{A},j^\text{B})},(\Delta^\text{A},\Delta^\text{B})} \simeq \ket*{\Phi^{(j^\text{A},j^\text{B})}(0,0)}$, the explicit form of the loaded state is given by
\begin{equation}
    \ket*{\Phi^{(j^\text{A},j^\text{B})}(0,0)} = \frac{r_\text{m}^\text{A} r^\text{opt,B} \ket{01} + (-1)^{j^\text{A}-j^\text{B}} r_\text{m}^\text{B} r^\text{opt, A} \ket{10}}{2\sqrt{2}}.
\end{equation}
When the condition $r^\text{opt,A}r_\text{m}^\text{B} = r_\text{m}^\text{A}r^\text{opt,B}$ is satisfied, which is identical to the condition for the sequential CAPS networking with single photons, the state reduces to the desired Bell state.
By adjusting the mirror reflectivities as specified in  Eq.~\eqref{app_eq:mirror_reflectivities_for_inhomogeneity}, unit fidelity is achieved in the long pulse limit, with a corresponding success probability of $[\min(r^\text{opt,A},r^\text{opt,B})]^2$.

\subsection{Emission-CAPS networking} \label{s_sec:type-III}
Emission-CAPS networking consists of an atom-photon entanglement generation followed by memory loading.
Alice first prepares the atom-photon Bell state,
\begin{equation}
    \ket{\Phi^{+}; {f}}_{ap} = \frac{\ket{0}_a^\text{A} \ket*{0; {f}}_p + \ket{1}_a^\text{A} \ket*{1; {f}}_p}{\sqrt{2}},
\end{equation}
which can be realized with, e.g., a four-level system inside a cavity (see Appendix~\ref{S_seq:atom_photon_entanglement}).
The photon is sent to Bob and loaded into the atomic qubit, ideally resulting in atom-atom Bell states.
According to the detailed analysis of the memory loading scheme in Appendix~\ref{s_sec:memory_loading}, the state of the two atomic qubits after the photonic qubit measurement with outcome $j \in \{0,1\}$ is given using the memory-loading operator $\hat{E}(\Delta)$ defined in Eq.~\eqref{S_eq:memoryloading}:
\begin{equation}
    \hat{\rho}^{(j)}_\text{ec} = \frac{1}{P^{(j)}_\text{ec}} \int\dd{\Delta}\frac{\vab{f(\Delta)}^2}{2} \hat{E}_a^\text{B}(\Delta) \ketbra*{\Phi_\text{id}^{(j)}}{\Phi_\text{id}^{(j)}} [\hat{E}_a^\text{B}(\Delta)]^\dagger,
\end{equation}
where
\begin{equation}
    P^{(j)}_\text{ec} = \int \dd{\Delta} \frac{\vab{f(\Delta)}^2}{2} \braket*[3]{\Phi_\text{id}^{(j)}}{[\hat{E}_a^\text{B}(\Delta)]^\dagger\hat{E}_a^\text{B}(\Delta)}{\Phi_\text{id}^{(j)}}
\end{equation}
represents the detection probability, and $\ket*{\Phi_\text{id}^{(0)}} = \ket*{\Phi^-}$ and $\ket*{\Phi_\text{id}^{(1)}} = \ket*{\Phi^+}$.
Thus, the total success probability and the conditional fidelity are respectively given by
\begin{equation} \label{app_ex:type-III_F_and_P}
    \begin{aligned}
        P_\text{ec} =& \int \dd{\Delta} \frac{\vab{f(\Delta)}^2}{2} \sum_{j=0,1} \braket*[3]{\Phi_\text{id}^{(j)}}{[\hat{E}_a^\text{B}(\Delta)]^\dagger\hat{E}_a^\text{B}(\Delta)}{\Phi_\text{id}^{(j)}}, \\
        F_\text{ec} =& \frac{1}{P_\text{ec}} \int \dd{\Delta} \frac{\vab{f(\Delta)}^2}{2}  \sum_{j=0,1} |\braket*[3]{\Phi_\text{id}^{(j)}}{\hat{E}_a^\text{B}(\Delta)}{\Phi_\text{id}^{(j)}}|^2.
    \end{aligned}
\end{equation}

\subsubsection{Photon in a mixed state}
As in Appendix~\ref{s_sec:type-II-networking_with_photon}, we again consider the case where the photon is generated in a mixed state.
For simplicity, we model such an atom-photon state as follows:
\begin{equation} \label{app_eq:atom_photon_Bell_state}
    \hat{\rho}_{ap} = \sum_l p_l \ketbra*{\Phi^{+}; {u}_l}[_{ap}]{\Phi^{+}; {u}_l} + \ab(1-\sum_l p_l) \hat{\rho}_{a\varnothing},
\end{equation}
where $\hat{\rho}_{a\varnothing}$ represents the state with the photonic state in $\ket{\varnothing}_p$.
As in the sequential CAPS networking, we straightforwardly obtain the fidelity and the success probability by replacing $|f(\Delta)|^2$ with $\sum_l p_l |u_l(\Delta)|^2$ in Eq.~\eqref{app_ex:type-III_F_and_P}.

\subsection{HEG with two-photon interference} \label{s_sec:type-I}
To clarify how the photon purity affects the generated Bell states in the two-photon-interference-based protocol, we present the fidelity of the atom-photon Bell states given by Eq.~\eqref{app_eq:atom_photon_Bell_state}.
For the case of polarization encoding used for the photonic qubit, the four detection patterns announce the generation of the remote Bell state with the same fidelity and success probability.
Here, we consider one of them, for which the measurement operator is given by~\cite{Kikura2025}
\begin{equation}
    \begin{aligned}
        \hat{\mathcal{D}}_\text{I}(t_0, t_1) =& \hat{\mathcal{P}}_\text{I}^\dagger(t_0, t_1) \hat{\mathcal{P}}_\text{I}(t_0, t_1), \\
        \hat{\mathcal{P}}_\text{I}(t_0, t_1) =& \hat{a}_0^{+}(t_0)\hat{a}_1^{+}(t_1),
    \end{aligned}
\end{equation}
where $\hat{a}_j^{\pm}(t) = [\hat{a}_j^\text{A}(t)\pm \hat{a}_j^\text{B}(t)]/\sqrt{2}$, and $t_j$ denotes the detection time of the photon $j$.
For the initial state $\hat{\rho}_{ap}^\text{A} \otimes \hat{\rho}_{ap}^\text{B}$, the atom-atom state after the measurement is given by
\begin{equation}
    \hat{\rho}_\text{I}(t_0,t_1) = \frac{\Tr_p[\hat{\mathcal{D}}_\text{I}(t_0, t_1) \hat{\rho}_{ap}^\text{A} \otimes \hat{\rho}_{ap}^\text{B}]}{\Tr[\hat{\mathcal{D}}_\text{I}(t_0, t_1) \hat{\rho}_{ap}^\text{A} \otimes \hat{\rho}_{ap}^\text{B}]},
\end{equation}
along with the probability density $p(t_0,t_1) = \Tr[\hat{\mathcal{D}}_\text{I}(t_0, t_1) \hat{\rho}_{ap}^\text{A} \otimes \hat{\rho}_{ap}^\text{B}]$,
where $\Tr_p[\cdot]$ represents the partial trace of the photonic state.
From the relation:
\begin{equation}
    \begin{aligned}
        &\hat{\mathcal{P}}_\text{I}(t_0, t_1)  \ket*{\Phi^{+}; {u}_l^\text{A}}_{ap}^\text{A} \ket*{\Phi^{+}; {u}_{l^\prime}^\text{B}}_{ap}^\text{B} \\
        &= \frac{1}{4} [u_l^\text{A}(t_0) u_{l^\prime}^\text{B}(t_1)\ket{01} + u_l^\text{A}(t_1) u_{l^\prime}^\text{B}(t_0)\ket{10}]\ket{\varnothing}^\text{A}\ket{\varnothing}^\text{B},
    \end{aligned}
\end{equation}
we find
\begin{widetext}
    \begin{equation}
        \hat{\rho}_\text{I}(t_0,t_1) = \frac{1}{16 p(t_0,t_1)}\mqty{g^{(1)\text{A}}(t_0,t_0)g^{(1)\text{B}}(t_1,t_1) & [g^{(1)\text{A}}(t_0,t_1)]^\ast g^{(1)\text{B}}(t_0,t_1) \\ g^{(1)\text{A}}(t_0,t_1) [g^{(1)\text{B}}(t_0,t_1)]^\ast & g^{(1)\text{A}}(t_1,t_1)g^{(1)\text{B}}(t_0,t_0)},
    \end{equation}
\end{widetext}
and
\begin{equation}
    \begin{aligned}
        &p(t_0,t_1) \\
        &= \frac{g^{(1)\text{A}}(t_0,t_0)g^{(1)\text{B}}(t_1,t_1)+g^{(1)\text{A}}(t_1,t_1)g^{(1)\text{B}}(t_0,t_0)}{16},
    \end{aligned}
\end{equation}
where the basis of the matrix is $\{\ket{01}, \ket{10}\}$.
Thus, the fidelity to the desired Bell state $\ket{\Psi^{+}}$ is given by
\begin{equation}
    F_\text{ee}(t_0,t_1) = \frac{1 + M^\text{AB}(t_0,t_1)}{2},
\end{equation}
where
\begin{equation}
    M^\text{AB}(t_0,t_1) = \frac{\Re\ab[[g^{(1)\text{A}}(t_0,t_1)]^\ast g^{(1)\text{B}}(t_0,t_1)]}{8p(t_0,t_1)},
\end{equation}
thereby resulting in the average conditional fidelity given by
\begin{equation}
    \begin{aligned}
        F_\text{ee} = \frac{\iint \mathrm{d}t_0 \dd{t_1} p(t_0,t_1)F_\text{I}(t_0,t_1)}{\iint \mathrm{d}t_0 \dd{t_1} p(t_0,t_1)} = \frac{1+M^\text{AB}}{2},
    \end{aligned}
\end{equation}
where
\begin{equation}
    M^\text{AB} = \frac{\iint \mathrm{d}t_0 \dd{t_1} \Re\ab[[g^{(1)\text{A}}(t_0,t_1)]^\ast g^{(1)\text{B}}(t_0,t_1)]}{\ab[\int \dd{t} g^{(1)\text{A}}(t,t)] \ab[\int \dd{t} g^{(1)\text{B}}(t,t)]},
\end{equation}
which is known as a mean-wavepacket overlap~\cite{Ollivier2021}.

For the two identical systems, $g^{(1)\text{A}}(t_0,t_1) = g^{(1)\text{B}}(t_0,t_1) [=g^{(1)}(t_0,t_1)]$, this reduces to
\begin{equation}
    F_\text{ee}  = \frac{1+M_\text{s}}{2},
\end{equation}
where $M_\text{s}$ is a single-photon trace purity~\cite{Fischer2018, Trivedi2020, Kikura2025_high_purity},
\begin{equation}
    M_\text{s} = \frac{\iint \mathrm{d}t\dd{t^\prime} |g^{(1)}(t,t^\prime)|^2}{\ab[\int \dd{t} g^{(1)}(t,t)]^2} = \frac{\sum_k \lambda_k^2}{(\sum_k \lambda_k)^2},
\end{equation}
which can be evaluated by a Hong-Ou-Mandel visibility~\cite{Ollivier2021}.
Note that a similar result has been derived in Ref.~\cite{Craddock2019}.

\section{Crosstalk in multi-atom CAPS gates} \label{s_sec:tbm_crosstalk_effect}
Here, we address the crosstalk effects that are critical for the fidelity of time-multiplexed CAPS gate operations.
In this protocol, a single target atom undergoes the CAPS gate interaction while the remaining $N_a - 1$ atoms are spectrally decoupled from the cavity via large ac Stark shifts.
Despite this detuning, the collective coupling of these spectator atoms to the cavity mode can still induce residual interactions that affect the gate fidelity of the target atom.
To quantitatively evaluate this effect, we develop a theoretical framework that allows us to derive an analytic expression for the crosstalk-induced infidelity, revealing its scaling with key parameters such as the detuning $\Delta_a$, atom number $N_a$, and internal cooperativity $C_\mathrm{in}$.
We outline the derivation of this analytical result below.

As a starting point, we extend the single-atom CAPS gate analysis to the case where $N_a$ atoms are confined within a single cavity.
For simplicity, we designate the atom with index $j=1$ as the target, and define the corresponding unitary operator as
\begin{equation}
    \begin{aligned}
        \hat{U}_\text{tar}^{(N_a)} =& \bm{1}_a^{\otimes N_a} \otimes \ketbra{0}[_p]{0}  \\
        & + (-\ketbra{0}[_{a}]{0}+\ketbra{1}[_{a}]{1}) \otimes \bm{1}_a^{\otimes N_a-1} \otimes \ketbra{1}[_p]{1}.
    \end{aligned}
\end{equation}
The corresponding Kraus operator $\hat{G}_0^{(N_a)}$ for $N_a$ atoms is given by
\begin{equation}
    \begin{aligned}
        \hat{G}_0^{(N_a)} =& r_{\text{m}} \bm{1}_a^{\otimes N_a} \otimes \ketbra{0}[_p]{0} \\
        &+ \sum_{\bm{j}[1;N_a]}r_{\bm{j}[1;N_a]} \ketbra{\bm{j}[1;N_a]}[_a]{\bm{j}[1;N_a]} \otimes \ketbra{1}[_p]{1},
    \end{aligned}
\end{equation}
where $\bm{j}[k;k^\prime]$ represents the bit string $j_kj_{k+1}\cdots j_{k^\prime}$ , and $\ket{\bm{j}[k;k^\prime]}_a = \ket{j_k}_{a}\ket{j_{k+1}}_{a}\cdots \ket{j_{k^\prime}}_{a}$.
Thus, we find
\begin{equation}
    \begin{aligned}\label{S_eq:Fpro-Leak}
        L^{(N_a)} =& 1-\frac{2^{N_a}|r_{\text{m}}|^2 + \sum_{\bm{j}[1;N_a]}|r_{\bm{j}[1;N_a]}|^2}{d_\text{q}}, \\
        F_\text{pro}^{(N_a)} =& \frac{|2^{N_a}r_{\text{m}} + \sum_{\bm{j}[2;N_a]}(-r_{0\bm{j}[2:N_a]} + r_{1\bm{j}[2:N_a]})|^2}{d_\text{q}^2},
    \end{aligned}
\end{equation}
with $d_\text{q} = 2^{N_a+1}$, leading to the conditional infidelity as
\begin{equation} \label{app_eq:def_of_conditional_infid_of_N_a_atoms}
    1- {F}_c^{(N_a)} = \frac{d_\text{q}}{d_\text{q}+1}\ab[1-\frac{F_\text{pro}^{(N_a)}}{1-L^{(N_a)}}].
\end{equation}

Next, we explicitly compute the conditional infidelity of Eq.~(\ref{app_eq:def_of_conditional_infid_of_N_a_atoms}) using the state-dependent reflectivity of the atom–cavity system.
We consider the case where atoms $j_2,j_3,\cdots j_N$ are detuned from the cavity resonant by an amount $\Delta_a$, which leads to the following modified reflection coefficients:
\begin{equation}
    \begin{aligned}
        r_{0\bm{j}[2;N_a]} =& 1- 2\eta\ab(1 + \frac{2mC}{1+ i\Delta_a/\gamma})^{-1} [\eqqcolon r_0^{(m)}], \\
        r_{1\bm{j}[2;N_a]} =& 1- 2\eta\ab(1 + 2C +  \frac{2mC}{1 + i\Delta_a/\gamma})^{-1} [\eqqcolon r_1^{(m)}],
    \end{aligned}
\end{equation}
where $\eta = \kappa_\text{ex}/\kappa$ and $C=g^2/(2\kappa\gamma)$.
Here, $m = \sum_{k = 2}^{N} j_k$ denotes the number of atoms $j_2,j_3,\cdots, j_{N_a}$ in $\ket{1}_a$.
To proceed, we evaluate the conditional infidelity in the regime where $|\Delta_a|/\gamma \gg N_a C, 1$, allowing us to neglect third- and higher-order terms in the small parameter $\epsilon = g^2/\kappa \Delta_a = 2C\gamma/\Delta_a$.
Since we are interested in the parameter regime with $C>1$, we also omit terms of $\mathcal{O}(\epsilon \gamma/\Delta_a)$, which contribute negligibly under these conditions.
In this regime, we find
\begin{equation}
    \begin{aligned}
        &\ab(1 + 2C +  \frac{2mC}{1 + i\Delta_a/\gamma})^{-1} \\
        &\simeq \frac{1}{1+2C} \ab[1+ m \frac{i\epsilon}{1+2C} - m^2 \ab(\frac{\epsilon}{1+2C})^2],
    \end{aligned}
\end{equation}
leading to the approximate expressions
\begin{equation} \label{S_eq:r_j^m_approximated}
    \begin{aligned}
        r_0^{(m)} \simeq& r_0 -2\eta (im\epsilon - m^2 \epsilon^2), \\
        r_1^{(m)} \simeq& r_1 -2\eta \ab[ \frac{im\epsilon}{(1+2C)^2} - \frac{m^2\epsilon^2}{(1+2C)^3}],
    \end{aligned}
\end{equation}
where the on-resonant single-atom reflectivities are given by
\begin{equation}
    r_0 = 1-2\eta, \quad r_1 = 1-\frac{2\eta}{1+2C}.
\end{equation}
By using Eq.~(\ref{S_eq:r_j^m_approximated}), we explicitly evaluate Eq.~(\ref{S_eq:Fpro-Leak}) under the conditions of both reflectivity and temporal-mode matching, $-r_0 = r_1 = r_\text{m} = r^{\text{opt}}$, given in Eq.~\eqref{eq:opt_r}.
In the following, we also assume $N_a\gg 1$ to simplify the expression, leading to,
\begin{equation}
    \begin{aligned}
        F_\text{pro}^{(N_a)} \simeq&  (r^\text{opt})^2\ab[1-\frac{1}{4}\frac{(r^\text{opt})^2 + 2}{(1+r^\text{opt})^2}(N_a\epsilon)^2], \\
        1-L^{(N_a)} \simeq&  (r^\text{opt})^2 \ab[1 + \frac{1}{8}\frac{1-r^\text{opt}}{1+r^\text{opt}}\frac{1+(r^\text{opt})^2}{(r^\text{opt})^2}(N_a\epsilon)^2].
    \end{aligned}
\end{equation}
Finally, we obtain the conditional fidelity and success probability at $C_\mathrm{in}\gg1$
\begin{equation}
    \begin{aligned}
    1- {F}_c^{(N_a)} &\approx \frac{1}{2}\ab(1+\frac{3}{4}C_\text{in})\ab(\frac{N_a \gamma}{\Delta_a})^2,\\
    {P}_\text{CAPS}^{(N_a)} &\approx (r^\text{opt})^2.
    \end{aligned}
\end{equation}

\section{Modeling wavelength-multiplexed CAPS gates} \label{s_sec:transfer-matrix_approach}
\begin{figure}
    \centering
    \includegraphics[width=\linewidth]{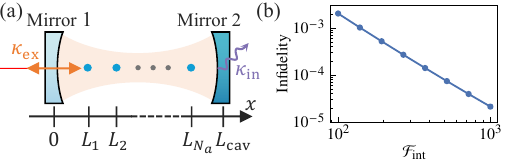}
    \caption{(a) Schematic of multiple atoms coupled to a cavity. For $N_a$ atoms within a cavity, $L_j~(j=1,2, \cdots, N_a)$ represents the position of the atom $j$.
    (b) Average infidelity as a function of the intrinsic finesse $\mathcal{F}_\mathrm{int}$ with $N_a = 5$, where the parameters are $\sigma_0/A_\text{eff} = 0.1, c/v_g = 1.4$, and $\gamma/2\pi = \qty{0.24}{\MHz}$.}
    \label{fig10}
\end{figure}

Wavelength-multiplexed CAPS operation requires the use of multiple cavity modes spaced by the free spectral range.
In this regime, the standard single-mode approximation---such as the frequency-dependent reflection model used in Eq.~\eqref{app_eq:freq-dep-reflection}---is no longer valid, as it neglects contributions from adjacent resonant modes.
To capture the effects of multiple cavity resonances, we adopt a transfer-matrix method---a practical framework for modeling the optical response of multi-atom, multi-mode cavity-QED systems.
This approach assumes a linear optical response, which is well justified for the CAPS gate operating with a single incident photon interacting with one atom at a time.

In the following, we implement the transfer-matrix method~\cite{Deutsch1995}, in which each component---such as atoms $\bm{M}_a$, mirrors $\bm{M}_{m1(2)}$, and propagation segments $\bm{M}_p$---is represented by a $2 \times 2$ matrix.
The application of this method to cavity-QED systems has been studied in detail in Ref.~\cite{Nemet2020}.
The overall transfer matrix of the system is constructed as the ordered product of these component matrices:
\begin{equation}
    \bm{M}_\text{cav} = \bm{M}_{m1} M_p(\Delta L_{0}) \ab[\prod_{j=1}^{N} \bm{M}_{aj} \bm{M}_{p}(\Delta L_{j})]\bm{M}_{m2},
\end{equation}
where $\Delta L_j = L_{j+1}- L_{j}~(L_0 = 0, L_{N_a+1} = L_\text{cav})$ [Fig.~\ref{fig10}(a)], and each matrix is explained in the following.
The reflection coefficient $r_\text{cav}$ of the system is given by
\begin{equation}
    r_\text{cav} = \frac{(\bm{M}_\text{cav})_{21}}{(\bm{M}_\text{cav})_{11}}.
\end{equation}

The matrix $M_{m1(2)}$ represents mirror $1(2)$ forming the cavity.
To employ the boundary condition being consistent with the conventional one in quantum optics~\cite{Raymer2024} and ensuring that the mirrors behave as fixed ends, we set the matrices as
\begin{equation}
    \begin{aligned}
        \bm{M}_{m1} =& \frac{1}{\sqrt{T_\text{ex}}}\mqty{1 & \sqrt{1-T_\text{ex}}\\\sqrt{1-T_\text{ex}} & 1}, \\
        \bm{M}_{m2} =& \frac{1}{\sqrt{T_\text{in}}}\mqty{1 & -\sqrt{1-T_\text{in}}\\\sqrt{1-T_\text{in}} & 1},
    \end{aligned}
\end{equation}
where $T_{\text{ex(in)}}$ denotes the transmittance of mirror 1(2).
Note that our definitions of mirror matrices differ from those adopted in Ref.~\cite{Nemet2020}.
For mirror 1, which acts as the coupler between the cavity and the output field, the transmittance is related to the coupling rate $\kappa_\text{ex}$ as $T_\text{ex} = 4\pi\kappa_\text{ex}/\omega_\text{FSR}$.
For brevity, we treat the internal loss as the nonzero transmittance of mirror 2, leading to $T_{\text{in}} = 4\pi \kappa_{\text{in}}/\omega_\text{FSR}$.

The matrix $M_p(x)$ represents the free propagation of light by distance $x$, which is given by
\begin{equation}
    \bm{M}_p(x) = \mqty{\exp\ab(-i\pi \frac{\Delta+\omega_0}{\omega_\text{FSR}}\frac{x}{L_\text{cav}})&0\\ 0 & \exp\ab(i\pi \frac{\Delta+\omega_0}{\omega_\text{FSR}}\frac{x}{L_\text{cav}}) }.
\end{equation}

Finally, $\bm{M}_{aj}$ represents the atom $j$ at position $L_j$.
To clarify the explicit form of that matrix,  we consider a single two-level ($\ket{1}_a,\ket{e}_a$) atom coupled to a one-dimensional waveguide.
Considering that an itinerant single photon interacts with the atom, the atom exhibits a linear response, where the reflection and transmission coefficients at frequency $\Delta + \omega_0$ are respectively given as follows:
\begin{equation}
    \begin{aligned}
        r_a =& -\frac{\Gamma_\text{1D}}{\Gamma_\text{1D} + \Gamma - 2i(\Delta-\Delta_a)}, \\
        t_a =& 1 - \frac{\Gamma_\text{1D}}{\Gamma_\text{1D} + \Gamma - 2i(\Delta-\Delta_a)},
    \end{aligned}
\end{equation}
which are derived by solving the (non-Hermitian) Schr\"{o}dinger equation, with neither a steady-state approximation nor a weak-excitation approximation~\cite{Liao2015}.
Here, $\Gamma_\text{1D}$ is the radiative energy decay rate into the target mode, and $\Gamma = 2\gamma$ is the atomic spontaneous energy decay rate.
We note that $|r_a|^2 + |t_a|^2 \leq 1$ due to the atomic spontaneous decay (the equality holds if and only if $\Gamma = 0)$.
The transfer matrix for the atomic linear response is given by~\cite{Deutsch1995}
\begin{equation} \label{S_eq:M_a}
    \bm{M}_a = \frac{1}{t_a}\mqty{1 & -r_a \\ r_a & t_a^2-r_a^2} = \mqty{1+i\zeta & i\zeta \\ -i\zeta& 1-i\zeta},
\end{equation}
where
\begin{equation}
    \zeta = \frac{\Gamma_\text{1D}}{2(\Delta-\Delta_a) + i\Gamma}.
\end{equation}
For the atom $j$, we set $\Delta_a$ to the detuning itself for $\ket{1}_a$, and to a sufficiently large value for $\ket{0}_a$.
The parameter $\Gamma_\text{1D}$ is related to the coupling strength $g$: $\Gamma_\text{1D} = \pi g^2/\omega_\text{FSR}$.

The transfer matrix approach yields the set of reflection coefficients $r_{\bm{j}[1; N_a]}$, which are used to calculate the fidelity for the target atom $j \in \{1, 2,\cdots,N_a\}$ by substituting them into Eq.~\eqref{app_eq:def_of_conditional_infid_of_N_a_atoms}.
We plot the average of values for each atom in Fig.~\ref{fig5}(c) and Fig.~\ref{fig10}(b).
Here, considering the position dependence of the coupling strength: $g(x) = g \sin(m \pi x/L_\text{cav})~(0 \leq x \leq L_\text{cav}, m \in\{1,2,\ldots\})$, the atoms are randomly placed at one of the antinodes within the region $0.45L_\text{cav} \leq x \leq 0.55L_\text{cav}$.
We evaluate 50 trials with different random configurations, where the external coupling rate $\kappa_\text{ex}$ is optimized for the unshifted target atom, and the plotted infidelity is averaged over atoms coupled to $N_\text{ch}$ distinct modes.

\bibliography{refs}
\end{document}